\documentclass[12pt,preprint]{aastex}
\usepackage{emulateapj5}

\shorttitle{X-ray Observation of the nuclei of FR I}
\shortauthors{Donato et al.}

\newcommand{\chandra}{{\it Chandra}} 
\newcommand{\xmm}{{\it XMM}} 
\newcommand{\asca}{{\it ASCA}} 
\newcommand{\hst}{{\it HST}}

\newcommand{\rosat}{{\it ROSAT}} 
\newcommand{\lum}{erg s$^{-1}$}
\newcommand{\flux}{erg cm$^{-2}$ s$^{-1}$}
\newcommand{\nh}{cm$^{-2}$}

\shorttitle{The X--ray Nuclei of FRIs}
\shortauthors{Donato et al.}

\def\ltsima{$\; \buildrel < \over \sim \;$}
\def\simlt{\lower.5ex\hbox{\ltsima}} % < over ~
\def\gtsima{$\; \buildrel > \over \sim \;$}
\def\simgt{\lower.5ex\hbox{\gtsima}} % > over ~

\begin{document}

\title{Obscuration and Origin of Nuclear X-ray emission \\
       in FR I Radio Galaxies}

\author{D. Donato}
\affil{George Mason University, School of Computational Sciences, 
              4400 University Drive, Fairfax, VA 22030}

\and

\author{R. M. Sambruna and M. Gliozzi}
\affil{George Mason University, Dept. of Physics \& Astronomy and 
School of Computational Sciences, MS 3F3,
             4400 University Drive, Fairfax, VA 22030}

\begin{abstract}
We present X--ray observations of the nuclear region of 25
Fanaroff-Riley I radio galaxies from the 3CRR and B2 catalogs, using
data from the \chandra\ and \xmm\ archives. We find the presence of a
X-ray Central Compact Core (CCCX) in 13/25 sources, in 3/25 sources
the detection of a CCCX is uncertain, while in the remaining 9/25
sources no CCCX is found. All the sources are embedded in a
diffuse soft X-ray component, generally on kpc-scales, which is in agreement
with the halo of the host galaxy and/or with the intracluster medium. 
The X-ray spectra of the cores are described by a power
law with photon indices $\Gamma=$1.1 - 2.6. In 8 sources excess
absorption over the Galactic value is detected, with rest-frame column
densities N$_H^z \sim 10^{20} - 10^{21}$ \nh; thus, we confirm the
previous claim based on optical data that most FRI radio galaxies lack
a standard optically-thick torus. We find significant correlations
between the X-ray core luminosity and the radio and optical
luminosities, suggesting that at least a fraction of the X-ray
emission originates in a jet; however, the origin of the X-rays 
remains ambiguous.
If the X-ray emission is entirely attributed to an isotropic,
accretion-related component, we find very small Eddington ratios,
$L_{bol}/L_{Edd} \sim 10^{-3}-10^{-8}$, and we calculate the radiative
efficiency to be $\eta \sim 10^{-2}-10^{-6}$, 
based on the Bondi accretion rates from the spatial analysis. This
suggests that radiatively inefficient accretion flows are present in
the cores of low-power radio galaxies. 
\end{abstract}


\section{Introduction}

According to unification schemes, radio galaxies are the mis-oriented
parent population of jet-dominated blazars, and thus correspond to
large viewing angles to the jet/torus axes. The low-power
(P$_{178~MHz} \lesssim 2 \times 10^{25}$ \lum), centrally-brightened
Fanaroff-Riley I (FRI) galaxies are the misaligned versions of BL Lac
objects, while high-power, edge-brightened FRIIs are unified with
powerful quasars.  Thus, if the unification models for radio-loud
sources hold, emission from the cores of both FRIs and FRIIs should be
significantly obscured by dust and gas contained by the torus.
Moreover, the accretion flow could be different at high and
low-powers, with FRIIs being powered by a standard disk and FRIs by an
ADAF \citep{Reyn, Ghis}.
These questions bear significance to the origin of the FRI/II dichotomy. 

While observations from optical to X-rays of FRIIs support the
presence of an obscuring dusty torus in these sources \citep{Samb, Ch00}, 
its presence in FRIs
is still highly controversial. Optical \hst\ images of FRIs show that
in most of these galaxies an unresolved nuclear point source is
present (Chiaberge, Capetti, \& Celotti 1999; Capetti et al. 2002, 
hereafter CH99 and CA02) arguing for lack of obscuration along the line of
sight. The optical flux correlates tightly with the radio core flux,
suggesting a non-thermal (synchrotron) origin of the optical emission
from the base of the jet (CH99). However, using the
same optical and radio data, \citet{Cao}
concluded that in most of the cases the emission from the central engine 
of FRIs is heavily obscured and is produced by a standard accretion disk.

At X-rays, FRIs were previously observed with \rosat\ and \asca. The
\rosat\ observations showed that the cores are embedded in diffuse
soft X-ray emission associated to the galaxy's ISM. The core X-ray
flux correlates with the radio flux, suggesting a jet origin for the
X-ray emission \citep{Hard00}. However, a correlation between X-ray
and radio fluxes may ensue from a more general correlation
(fundamental plane in 3-D space) between black hole mass, X-ray, and
radio emission, where the radio is produced by the jet whereas the
X-rays are related to the accretion flow \citep{Merl}. At energies
above 2 keV, the \asca\ spectra of 7 FRIs are best described by a
power-law component with photon index $\Gamma \sim 1.3-1.9$
\citep{Samb}. Due to the large PSF of the \asca\ detectors, however,
contributions from off-nuclear X-ray point sources cannot be excluded
(e.g., Turner et al. 1997). 

The combination of {\it XMM-Newton} (or simply \xmm) and \chandra,
with their complementary capabilities, is ideal to study these complex
X-ray sources and address, in particular, the issue of nuclear
obscuration.  The unprecedented spatial resolution of \chandra\ allows
one to disentangle the different components (unresolved core, kpc jet,
point like sources, diffuse emission) contributing to the X-ray
emission. Moreover, with \chandra\ one can investigate directly the
circumnuclear region, measuring the density and temperature profiles
close to the accretion radius of the central black hole.  On the other
hand, the larger effective area of \xmm\ and the combination of the
EPIC cameras working simultaneously results in a superior photon
yield, which allows more detailed analysis of the nuclear X-ray
spectrum.  A further advantage of \xmm\ is that also bright point-like
sources are not affected by pile-up problems, which is unfortunately
common in \chandra\ observations.  In fact, the advantage of the
complementary use of \xmm\ and \chandra\ was demonstrated in a 
previous study of the FRI 3C~270 \citep{Glio03}. 

Here we extend this study to archival FRIs in order to test the
unification models and the origin of the FRI/II division.  The basic
questions we will address are: i) Do FRIs have obscuring tori? Does
the column density correlate with luminosity, inclination angle? ii)
What is the origin of the nuclear X-ray emission? How much does the
beamed component contribute to the nuclear X-ray flux? If the X-rays
originate from the accretion flow, iii) What is the nature of
accretion in the nuclei of FRIs? 

The outline of the paper is as follows. In \S~2 we discuss the
sample selection criteria and in \S~3 the observations and data
analysis.  Results of the spatial and spectral analysis are given in
\S~4. Correlations between parameters are presented in \S~5, 
while the interpretation of the results is given in \S~6. Throughout
this paper, $H_0=75$ km s$^{-1}$ Mpc$^{-1}$ and $q_0=0.5$ are adopted.

\section{Sample and X-ray Observations}

We started with the samples presented in CH99 and CA02.
Chiaberge's sample contains 33 radio galaxies belonging to the 3CR catalog
\citep{Spin} and morphologically identified as FR~I
radio sources by \citet{Lain83} and/or \citet{Zirb}.  Capetti's
sample, on the other hands, contains 57 radio galaxies belonging to
the B2 catalog \citep{Coll, Fant}.  These samples contain all the {\it
bona fide} FRI radio galaxies observed in the optical with \hst. 

We cross-correlated the CH99 and CA02 samples with the \chandra\ and
\xmm\ archives, and selected all the sources with available X-ray
observations up to January 2004. The final sample, which contains 25
sources, is reported in Table~\ref{sources}. Also listed are
redshifts, radio, optical, and UV luminosities from CH99 and CA02. The
inclination angles of the radio jet with respect to the line of sight
in column 5 were obtained from the literature, listed in column
6. The inclination angles are usually determined from high resolution
radio data using the brightness ratio between the parsec-scale jet and
the counter-jet, and/or the arcsecond scale core dominance with respect
to the total power of the source (e.g., Giovannini et al. 2001). 
Unfortunately, for many sources no uncertainties are reported 
in the literature.
Column 7 lists the mass of the central black hole, derived from the
correlation between the stellar velocity dispersion of the host bulge
and its B-band magnitude \citep{Marc}.

Among the 25 sources available in the \chandra\ and \xmm\ archives,
11 have only \chandra\ observations (1 source with ACIS-I
and 10 with ACIS-S), 3 have only \xmm\ observations, and the
remaining 11 have been observed with both satellites.
For the sources with both \chandra\ and \xmm\ observations, the former
satellite has been used for the spatial analysis (except for 3C~84
which is affected by severe pile-up). Also, for the spectral analysis
\chandra\ observations have been preferred because of the ability to
disentangle the nuclear from the diffuse component; however,
in 4 cases (B2~0120+33, B2~0149+35, 3C~274.0, and 3C~338) \xmm\ 
observations have been used due to the poor \chandra\ spectra.
  
The log of the X-ray observations is reported in Table~\ref{log}.

\section{X-ray observations and data reduction}

\subsection{Chandra} 

The \chandra\ observations were carried out between 2000 April and
2004 January. All were performed with ACIS-S, with the sources at the
nominal aimpoint of the S3 chip, except for 3C~28 which was observed
with ACIS-I. In several cases, the original observers requested a
reduced frame time to mitigate core pile-up. In the case of 3C~78, no
reduced time frame was requested for the short
\chandra\ observations and the core suffers from significant pileup
(28\%). No spatial or spectral analysis was performed for this
object.  The ACIS event files were screened using \verb+CIAO+ 2.3
according to standard criteria.

Background spectra and light curves were extracted from source-free
regions on the same chip of the source. Time intervals corresponding
to background flares were excluded. The net
exposure times are listed in Table~\ref{log}. Spectra were extracted
from a circular region with radius 1.5\arcsec\ (see \S~4.1), and their
response matrices were constructed using the corresponding thread in
\verb+CIAO+ 2.3. The ACIS spectra were analyzed in the energy range
0.3--8 keV, where the calibration is best known and the background 
negligible.

\subsection{XMM-Newton}

Given that the cores of FRIs are relatively weak X-ray emitters, we
used only data from the EPIC pn and MOS cameras for our
analysis. These were performed using different observing modes
(extended and full frame) and different filters (thin and medium), as
specified by the original observers. 

No core pile-up was detected in either the pn or MOS cameras according
to the {\tt SAS} task {\tt epatplot}. The recorded events were
screened to remove known hot pixels and other data flagged as bad.
For \xmm, only data with {\tt FLAG=0} were used. The data were
processed using the latest CCD gain values, and only events
corresponding to pattern 0--12 (singles, doubles, triples, and
quadruples) in the MOS cameras and 0--4 (singles and doubles only,
since the pn pixels are larger) in the pn camera were accepted.  Arf
and rmf files were created with the \xmm\ Science Analysis Software
\verb+SAS+5.4.

EPIC spectra were extracted in circular regions centered on the core
and with radii 20\arcsec-30\arcsec, depending on the intensity and the
location on the chip of the source. Spectral analysis was performed in
the energy range 0.2--10 keV.

\section{Results}

\subsection{Spatial Analysis}

\begin{figure}
\begin{center}
\includegraphics{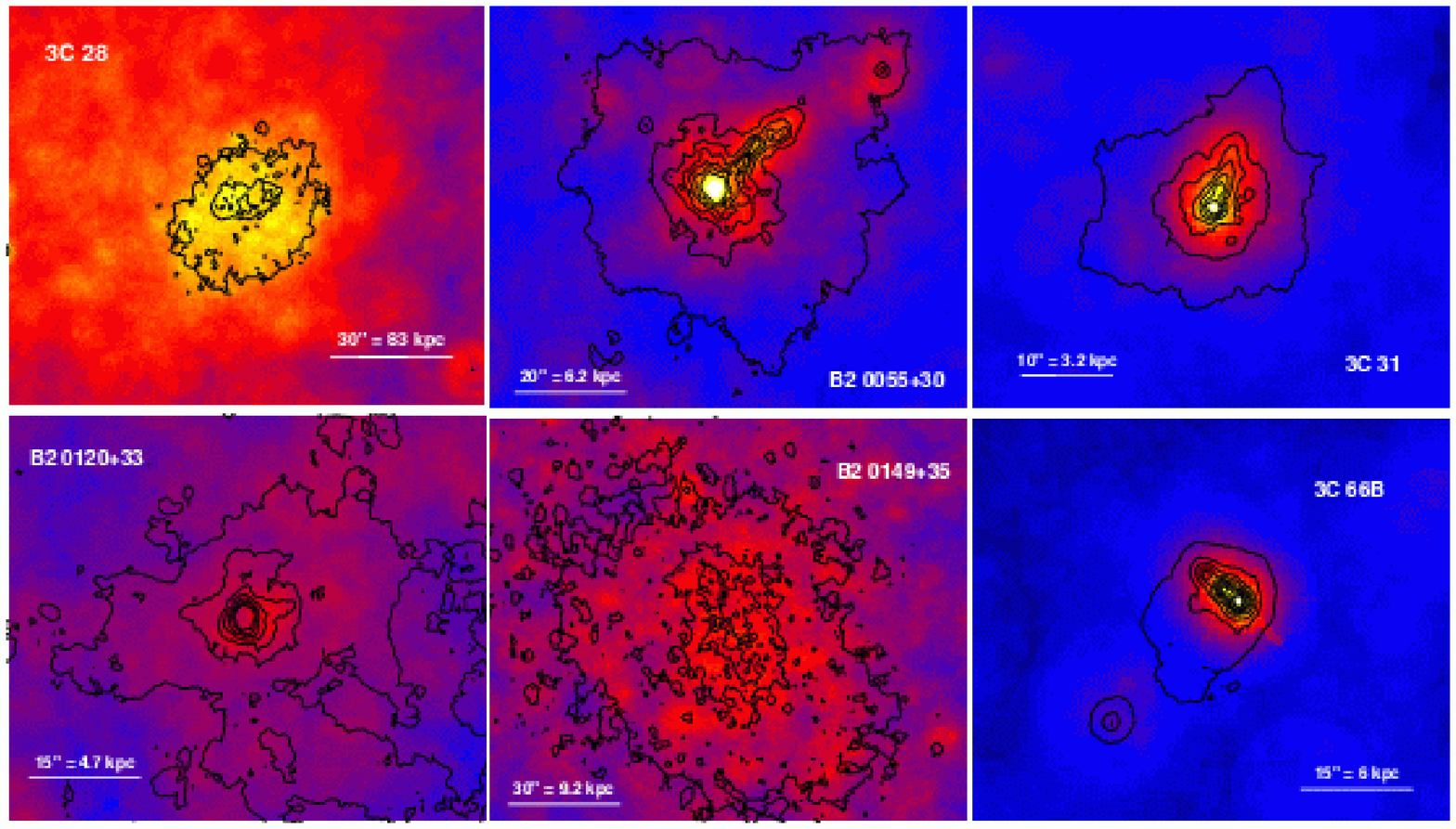}
\includegraphics{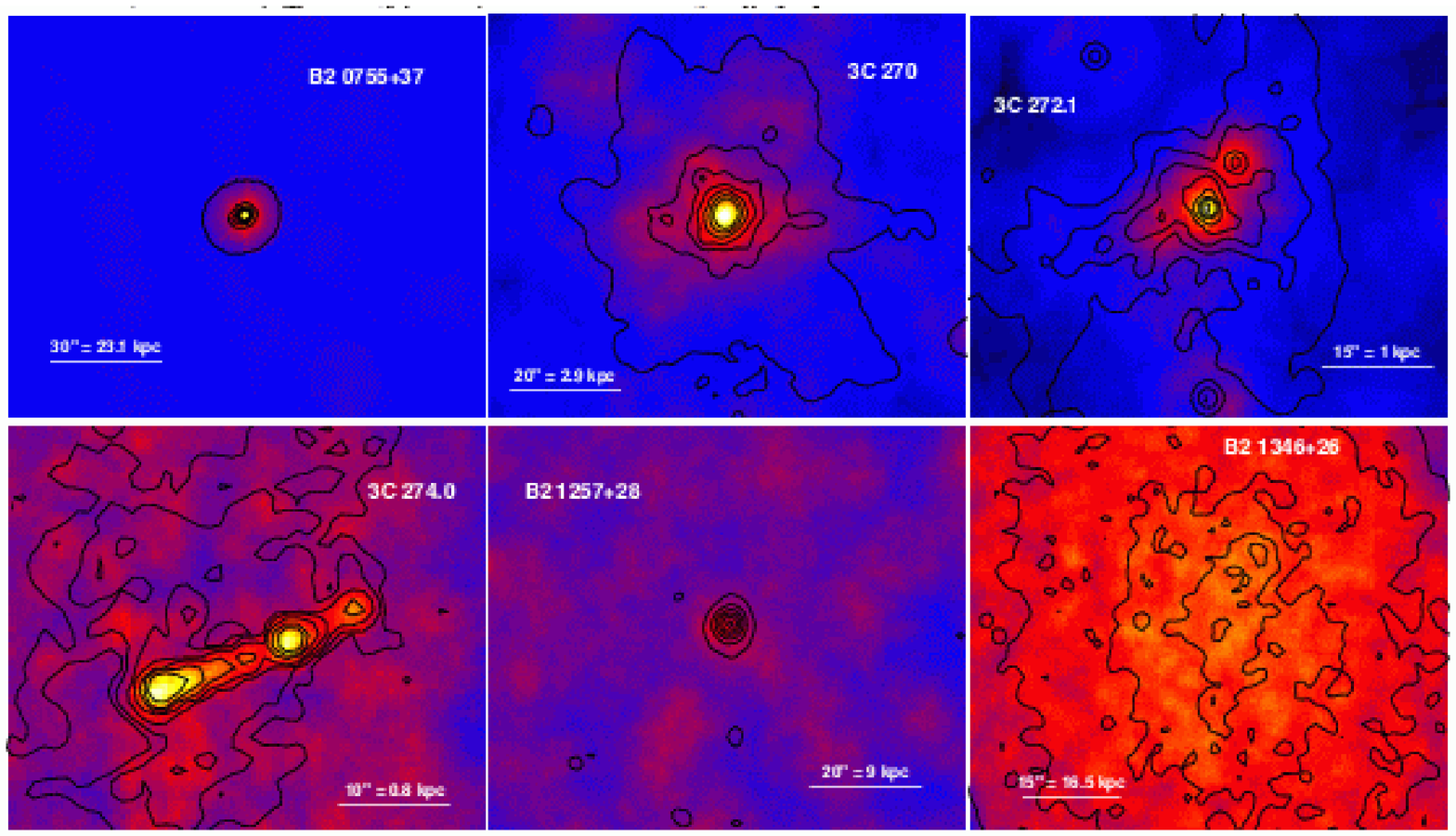}
\end{center}
\caption{Hard X-ray (2--8 keV) images of the FRI radio galaxies of our
sample, with the soft X-ray (0.3--2 keV) contours overlaid. All images
were obtained with the ACIS cameras on \chandra. 
The soft and hard X-ray images were smoothed using the sub-package 
{\tt fadapt} of {\tt FTOOLS} with a circular top hat
filter of adaptive size in order to achieve a minimal number of
10 counts under the filter.
A hard X-ray point source, embedded in
diffuse soft X-ray emission, is generally present (see Table~\ref{hardness}). 
}
\label{core1}
\end{figure}

\begin{figure}
\begin{center}
\includegraphics{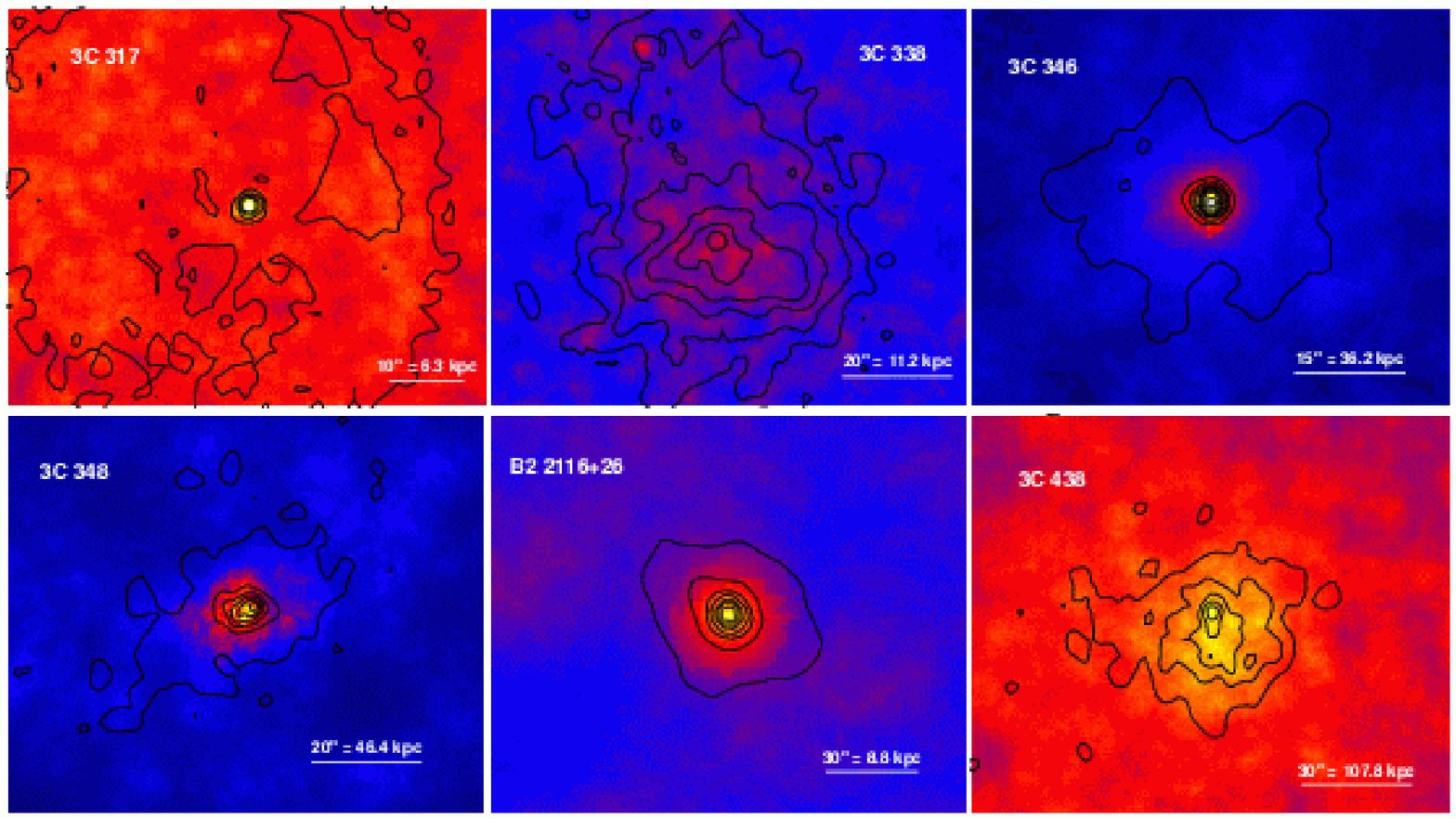}
\includegraphics{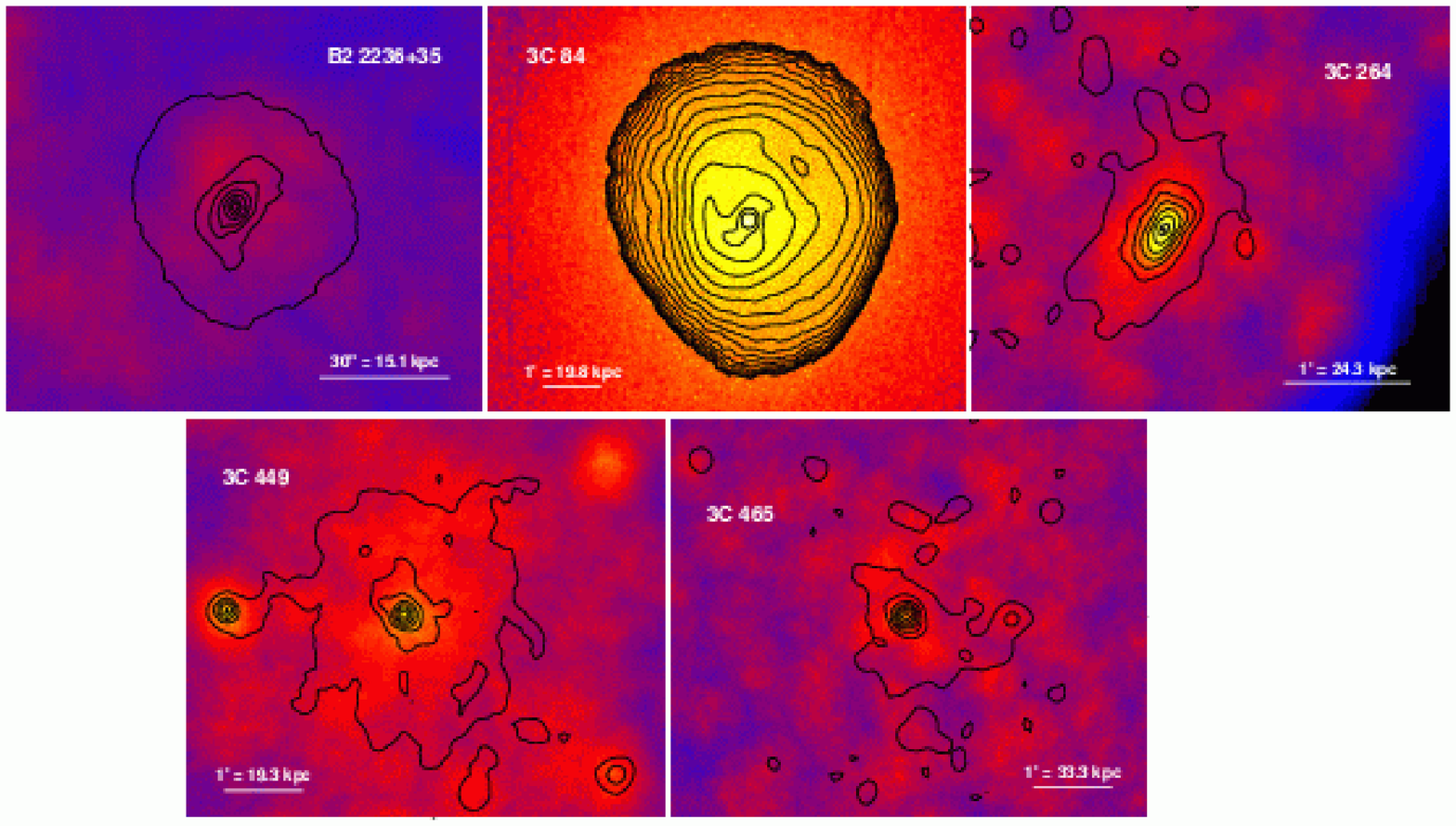}
\end{center}
\caption{Same as for Figure~~\ref{core1}, except that the X-ray images for 3C~84,
3C~264, 3C~449, and 3C~465 are from \xmm\ EPIC MOS1.}
\label{core2}
\end{figure}

Figure~\ref{core1} and \ref{core2} show the \chandra\ and \xmm\ 
images of the sources, with the
soft X-ray contours in 0.3--2 keV overlaid on the hard X-ray (2--8
keV) images. Both soft and hard X-ray images were smoothed 
using the sub-package {\it fadapt} of {\it FTOOLS} with a circular top hat
filter of adaptive size in order to achieve a minimal number of
10 counts under the filter.
Inspection of Figure~\ref{core1} and \ref{core2}  shows the presence of
diffuse soft X-ray emission in all sources. In most cases, the diffuse
emission is on kpc-scales and is associated with the host galaxy
halo. In 3C~28, B2~0120+33, B2~0149+35, 3C~84, 3C~272.1, B2~1346+26,
3C~317, and 3C~338 the soft X-rays appear to extend on
larger, cluster-like scales. Hard X-ray point sources are present in
several cases (see below). X-ray jets are present in 
B2~0055+30 \citep{Worr03}, 3C~31 \citep{Hard02}, 3c~66B \citep{Hard01}, 
B2~0755+37 \citep{Worr01}, 3C270 \citep{Ch03b}, and 3C274.0 
\citep{Mars,Wils}.

Since this paper focuses on the X-ray emission from the radio galaxy
cores, it is important to know in how many cases an unresolved X-ray
source is detected. To quantify this, we performed a detailed spatial
analysis of the \chandra\ and \xmm\ data. 

We adopted the following procedure. First, radial surface-brightness
profiles were extracted from a series of concentric annuli centered on
the radio core position. Off-nucleus X-ray point sources, as well as
the X-ray jet, were excluded. Second, the radial profiles were fitted
with a model including the instrument Point Spread Function (PSF) and
one or more $\beta$ models to describe the diffuse soft X-ray
emission. The significance of the PSF was determined using an F-test,
assuming as threshold for significant detection of the PSF a
probability P$_F=$99\%, corresponding to a 3$\sigma$ confidence
level. 

For the \xmm\ images, the PSF was described using the analytical
description of \citet{Ghiz}.  Since the aim-point of the EPIC-pn
camera if very close to the edge of the CCD (limiting the extension of
the extraction regions to few arcseconds), we decided to use only
EPIC-MOS1 data. For the \chandra\ images, the PSF was created using
the Chandra Ray Tracer ({\it ChaRT}) simulator which takes into
account the spectral distribution of the source. The ACIS PSF was thus
described using the 6-parameter function:

\begin{displaymath}
PSF(\rm{x})= A_0e^{-A_1{\rm{x}}^{A_2}}+A_3e^{-A_4{\rm{x}}^{A_5}}
\end{displaymath}

where the free parameters were determined by fitting the radial
profile of the PSF. 

The $\beta$ model is described by the following formula (e.g., 
Cavaliere \& Fusco-Femiano 1976):

\begin{displaymath}
S(r)=S_0\left(1+{r^2\over r_c^2}\right)^{-3\beta+1/2}.
\end{displaymath}

Table~\ref{hardness} summarizes the X-ray core detections, while
Table~\ref{beta} lists the best-fit parameters (core radius, $\beta$ 
value) for the $\beta$ models. Three examples of the fit of
radial profiles are shown in Figure~\ref{profili}.

Column 2 of Table~\ref{hardness} flags those FRIs where a Compact
Central Core (CCC) was detected with \hst\ in the optical. For
analogy, we will use the term CCCX to indicate the CCC counterpart in
the X-rays. As apparent from Table~\ref{hardness}, a CCCX was detected
in 13/25 sources, while no CCCX is present in 9 sources.  In the case
of B2~2116+26, the PSF is significant at P$_F$=95.6\%, corresponding
to a 2$\sigma$ confidence level. 
However, the \chandra\ observation is one of the shortest, with
only 9.6 ks of live time. We regard this source as a likely 
CCCX candidate.
The remaining 3 sources -3C~78, 3C~264, and 3C~272.1- are discussed
here individually.

3C~78: due to the strong core pileup, which distorts the shape of the
PSF, no reliable fit to the radial profiles could be performed. 
However, the presence of central pileup is {\it di per se} indication
of strong unresolved X-ray emission. Therefore, we conclude that a
CCCX is present in 3C~78. 

3C~264: the PSF is dramatically distorted due to the large offset from
nominal aimpoint of the source. Thus, no reliable spatial analysis can
be performed. However, the source spectrum (see Table~\ref{spec})
shows the presence of a hard X-ray power-law component. We interpret
this component as the signature of non-thermal emission from an AGN. 
We consider the detection of a CCCX uncertain in this case. 

3C~272.1: the \chandra\ image shows soft X-ray emission with a very
disturbed morphology. The radial profile cannot be properly fitted with
$\beta$-model(s) since the spatial fit does not converge. 
As the inspection of the hard X-ray image shows
the presence of a point source, conservatively we consider the CCCX
detection uncertain. 

Let us now compare the detection rate of the CCC and of the CCCX. We
find that: 

\begin{itemize} 

\item Of the 18 FRIs with a CCC in the optical, 13 have also a CCCX;  

\item All FRIs with a detected CCCX have a detected CCC, except 3C~438; 

\item No compact core was detected at either optical or X-rays in 
3 FRIs (3C~28, B2~0120+33, and B2~1257+28). 

\end{itemize} 

There are two possibilities for the non-detection of CCCX: either the
AGN X-ray emission is intrinsically weak (below the detection
threshold of \chandra\ or below the level of the circumnuclear 
extended emission), or it is absorbed by a very large amount of
cold gas, $N_H^z > 10^{24}$ \nh. 
The first case will be discussed in \S~4.3, while the second case
in \S~6.1.

\subsection{Variability Analysis} 

We searched for X-ray flux variability in the background-subtracted
light curves of the sources with positive CCCX detections. When
possible, \xmm\ data were used to take advantage of the larger
signal-to-noise ratio of the EPIC data. According to a $\chi^{2}$ test
for constancy, no significant variability of the 2--10 keV flux is
detected in the sources of our sample, except for 3C~270. For more
details on this source we refer to \citet{Glio03}. 

\subsection{Spectral Analysis}

The main goal of the X-ray spectral analysis is to investigate the 
physical conditions of the sources in the sample, focusing on the 
non-thermal emission from the 15 detected CCCX.
We also include the two CCCX candidates 3C~264
and 3C~272.1, while 3C~78 was not considered because of the strong
core pileup problems. 

The ACIS and EPIC spectra, extracted as
described above (\S~3), were grouped so that each new bin had \gtsima
20 counts to enable the use of the $\chi^{2}$ statistics. For 3C~438,
for which \ltsima 200 counts were detected, the X-ray
spectrum was not rebinned and the C-statistic was used instead.  The
spectra were fitted within
\verb+XSPEC+ v.11.2.0. Errors quoted throughout are 90\% for one
interesting parameter ($\Delta\chi^2$=2.7). 

The X-ray spectra were fitted with a two-component model, both
absorbed by Galactic $N_{\rm H}$.  At soft energies, the circumnuclear
emission from the host galaxy and/or from the cluster was
parameterized by a thermal component, the model
\verb+apec+ in \verb+XSPEC+, with temperature $kT$ and abundance
$Z/Z_{\odot}$. During the fits, the abundance was left free to vary
between 0.2 and 1, or otherwise fixed at one of these
two limits.  For sources embedded in a galaxy cluster, often more than
a single thermal component was requested to adequately fit the data,
since the cluster can have a gradient of temperatures and/or
abundances in the ACIS/EPIC extraction radii. 

At hard X-rays, a power-law model with photon index $\Gamma$ was used
to describe the CCCX non-thermal emission. The power law is absorbed
by a column density N$_H^{z}$ at the redshift of the source, thus
representing any excess intrinsic absorption over the Galactic
value. We note that most CCCX are at low $z$, thus absorption by the
ISM along the line of sight is likely not to be significant. The
significance of the power-law component over the thermal model was
determined with the F-test. 

\begin{figure}
\includegraphics[bb=75 40 485 618,clip=,width=5.5cm]{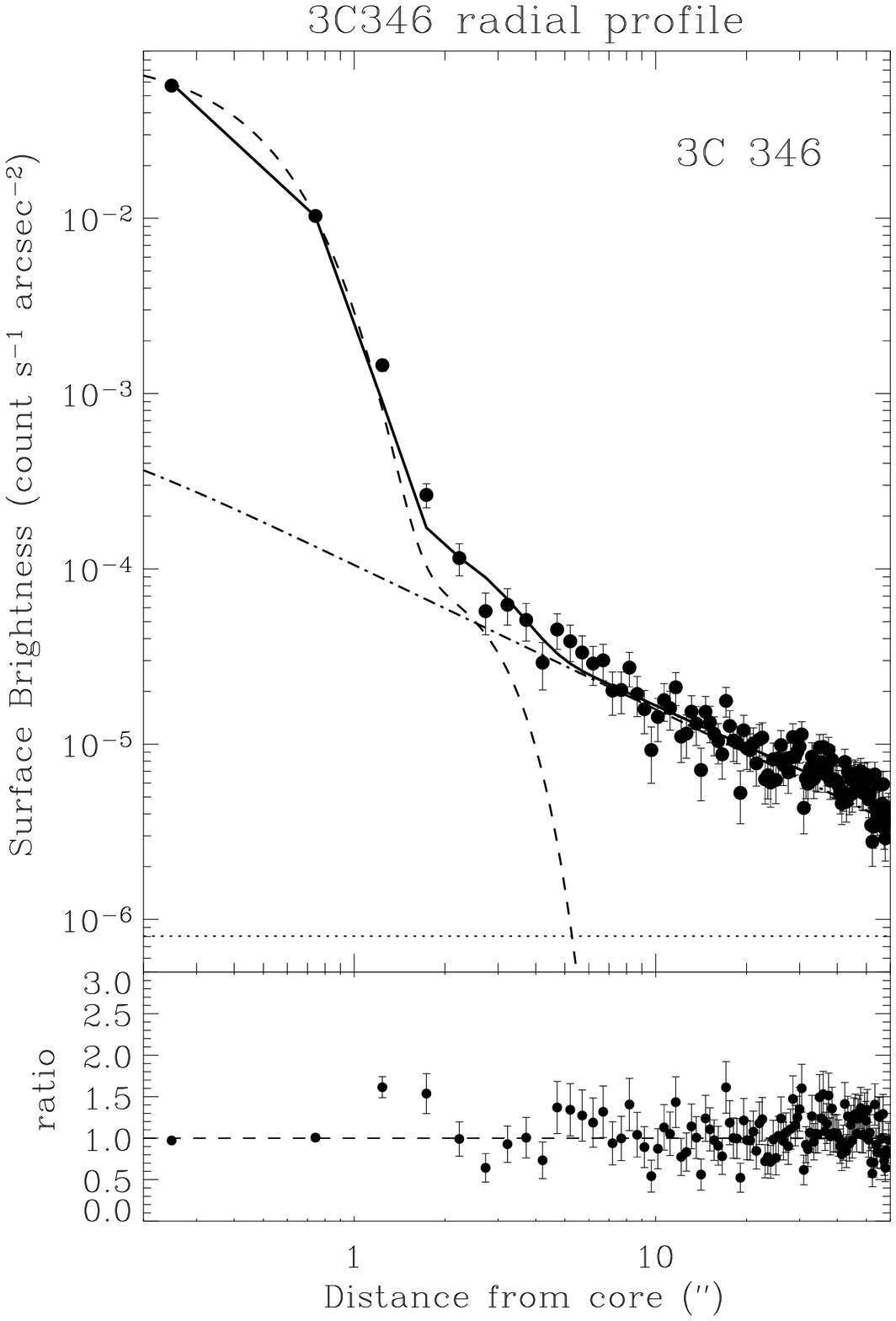}\includegraphics[bb=75 40 485 618,clip=,width=5.5cm]{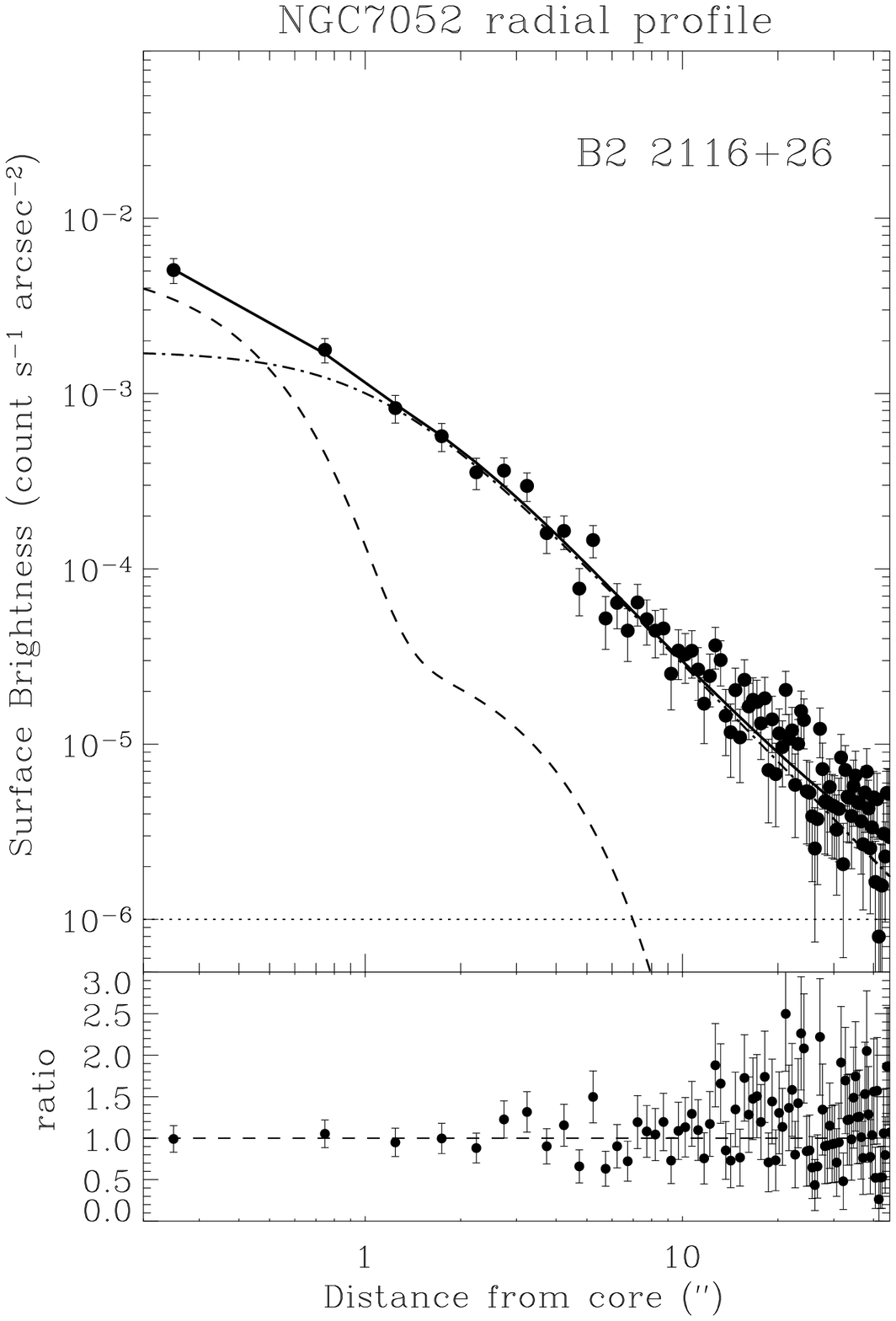}\includegraphics[bb=75 40 485 618,clip=,width=5.5cm]{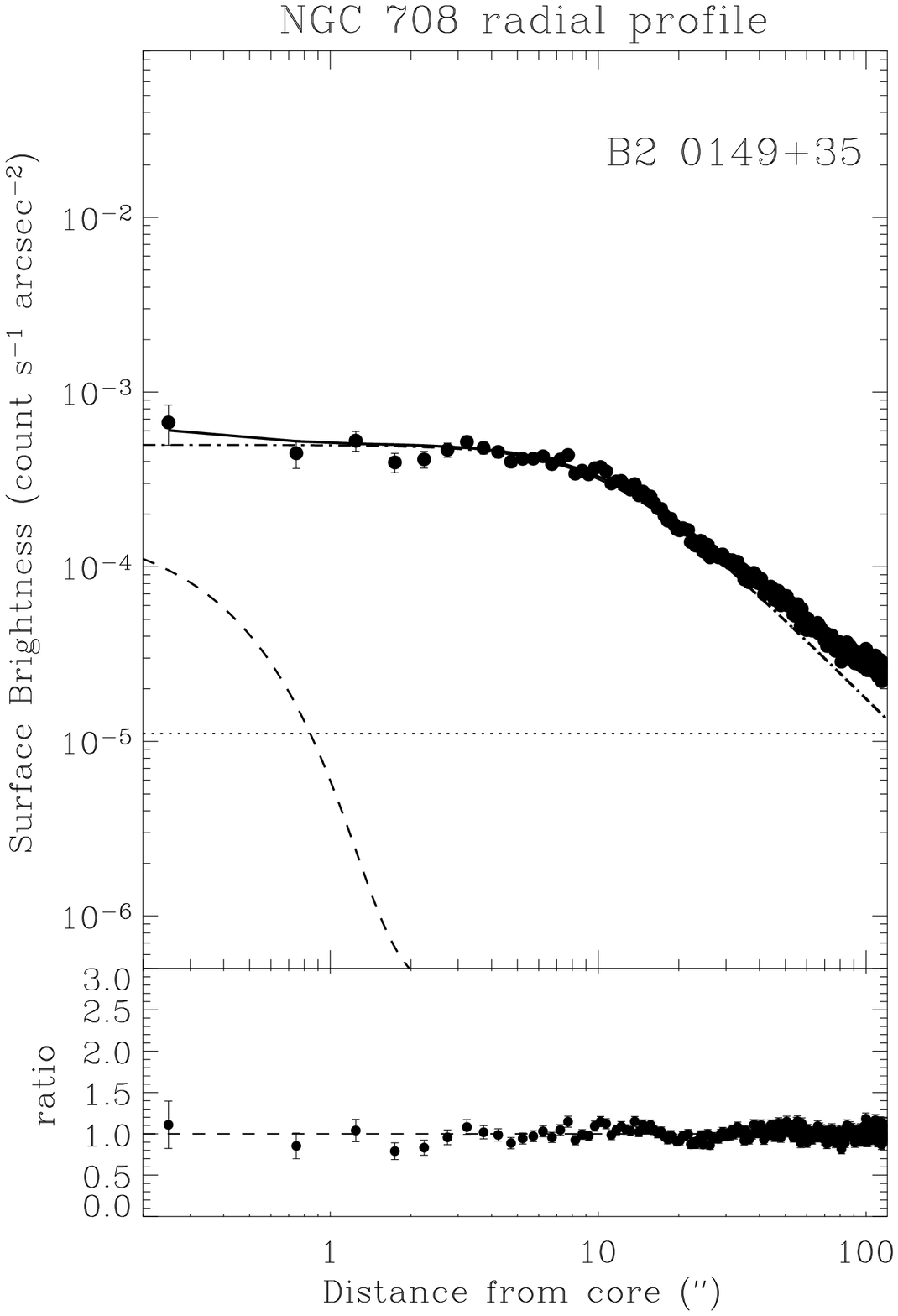}
\caption{Examples of radial profiles for three sources observed with
ACIS. In the top panels, the dashed line is the instrumental PSF, and
the dot-dashed line the $\beta$-model describing the diffuse
emission. The continuous thick line is the total model, while the thin
dotted horizontal line is the background. The residuals of the
best-fit model are shown in the bottom panels. The PSF is detected
with high ($>$ 99.9\%) significance in 3C~346, moderate ($\sim$ 96\%)
significance in B2~2116+26, while it is not requested in the case of
B2~0149+35.} 
\label{profili}
\end{figure}

\begin{figure}
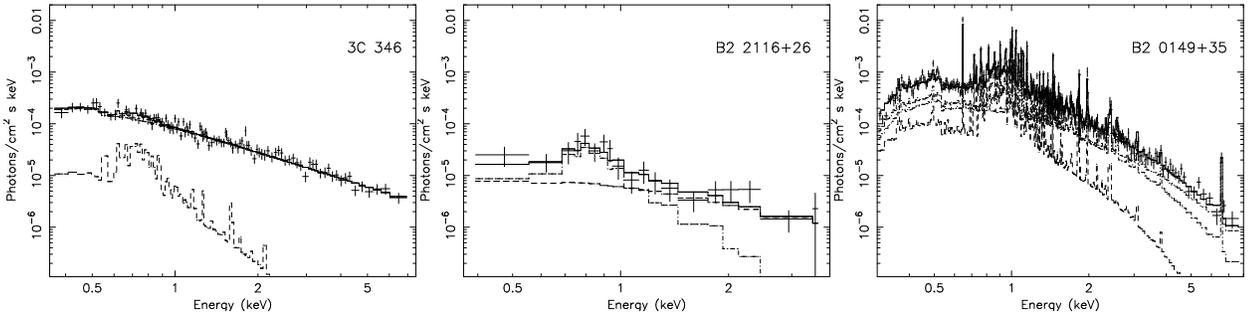

\includegraphics[angle=-90,width=5.5cm]{f27.eps}\includegraphics[angle=-90,width=5.5cm]{f28.eps}\includegraphics[angle=-90,width=5.5cm]{f29.eps}
\caption{X-ray spectra for the three sources presented in 
Figure~\ref{profili}. In agreement with the spatial analysis, a power-law
component is required for 3C~346, partially required for B2~2116+26, and 
no required for B2~0149+35. } 
\label{spectra}
\end{figure}

The results of the spectral fits are presented in
Table~\ref{spec}. Three spectra are shown in Figure~\ref{spectra}
corresponding to the 3 qualitatively different surface
brightness shown in Figure~\ref{profili}. For all the
sources with a detected CCC in the X-rays, the power-law component is
always required at high significance level (P$_F$ \gtsima 95\%). 
For the sources without a CCCX, the power law is not requested and 
the X-ray spectrum is adequately described by one or multiple thermal
components.

\begin{figure}
\centering
\includegraphics[width=8.0cm]{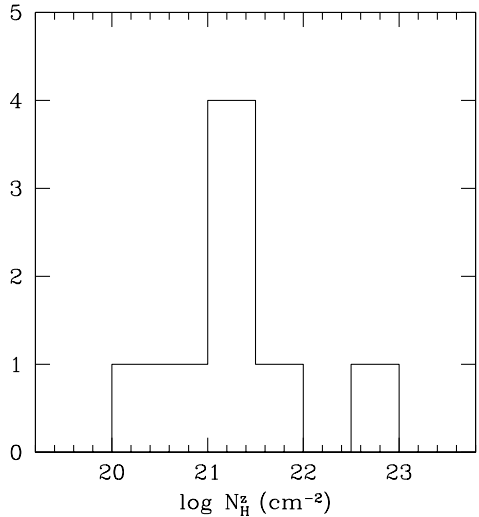}
\caption{Histogram of the intrinsic absorption obtained from the 
spectra of sources with detected CCCX.}
\label{istonh}
\end{figure}

In 8 sources, statistically significant absorption over the 
Galactic value is
detected, with N$_H^z \sim 10^{20} - 10^{21}$ \nh\ (see also 
Figure~\ref{istonh}). The source
3C~270 stands out for the largest intrinsic column density, N$_H^z
\sim 10^{22}$ \nh, in agreement with previous findings \citep{Glio03}. 
The power law photon index spans a wide range of values,
$\Gamma \sim 1.1 - 2.6$, with average value $\langle \Gamma \rangle =
1.9$ and standard deviation $\sigma=0.4$. 

Most of the FRIs of the our sample are embedded in diffuse emission on
the scale of the host galaxy halo, $\sim$ several kpc (Table 4). The
fitted temperatures are $kT \sim 0.3-1 keV$, in agreement with
previous results \citep{Samb, Worr00}. 

The observed fluxes and intrinsic (absorption-corrected) luminosities
in the energy range 0.3--8 keV are reported in Table~6. Also listed in
the Table are the observed fluxes and intrinsic luminosities in 0.3--8
keV for the power law component only. The latter span 3 orders of
magnitude, with L$_X \sim 10^{40} - 10^{43}$ \lum. Comparing the
values in Table~6, it is apparent that the AGN power-law 
emission typically contributes \gtsima 50\% of the total X-ray emission. 
This result is independently confirmed by the analysis of the radial
profiles: 
evaluating the integrated area under the PSF and under the 
$\beta$-model over the inner 20\arcsec\, 
we find that the PSF-to-total flux ratio is \gtsima 50\%. 

In the cases of 3C~28, B2~0149+35, and B2~2116+26, no power-law
component was required in the X-ray spectrum, in line with the fact that 
a CCCX is not detected in these sources. 
However, limits to the contribution of the X-ray emission due to 
the AGN can be derived from the radial profiles (e.g., Figure~\ref{profili}). 
While the PSF is not statistically significant, 
the normalization on the PSF can be used to calculate an upper
limit on the relative AGN contribution to the total
X-ray emission.
We find that the AGN contributions are of the order 
of $\sim$6\%, $\sim$1\%, and $\sim$30\%, respectively.

For the remaining 5 sources with undetected CCCX, we used
the most conservative value of the ratio ($\sim$1\%) to derive the
upper limit on the AGN luminosity.
The values are reported in part b) of Table~\ref{flux}.

\section{Correlations}

We have investigated possible trends among various parameters related
to the core emission, namely, the X-ray, optical, and radio
luminosities, the absorption column density, and the inclination angle
of the radio jet. The goal is to uncover clues on the origin of the
X-ray emission and on the presence of an obscuring torus, as expected
in the context of unification models. 

To quantify the degree of linear correlation, we calculated the linear
correlation coefficient $|r|$ and computed the chance probability
$P_{\rm r}(N)$ that a random sample of $N$ uncorrelated pairs of
measurements would yield a linear correlation coefficient equal or
larger than $|r|$. If the chance probability is small, the two
quantities are likely to be correlated. We use as minimum probability
of correlation $P_{\rm r}(N) <$ 1\% (which corresponds to a 3$\sigma$
level). To account for upper limits, we
used the generalized Kendall's Tau test contained in the statistical
package \verb+ASURV+ \citep{Lava}. The chance probabilities and the
linear correlation coefficients with the intercept coefficient and
slope of the linear regression (y=a+bx) are shown in Table~\ref{corr}a
for the detections and in Table~\ref{corr}b when limits are included.

\begin{figure}
\centering
\includegraphics[bb=40 175 550 675,width=8.0cm]{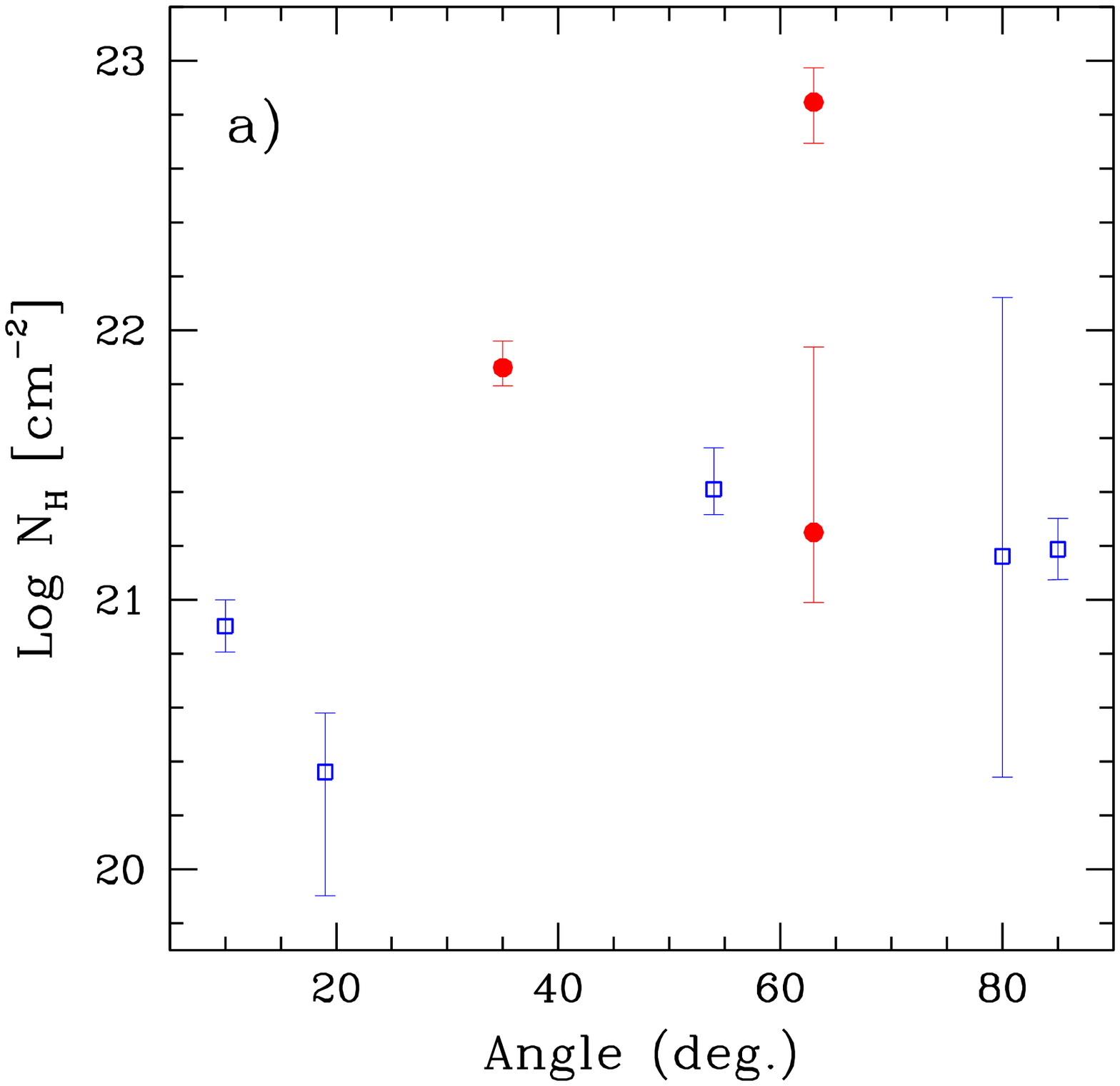}\includegraphics[bb=40 175 550 675,width=8.0cm]{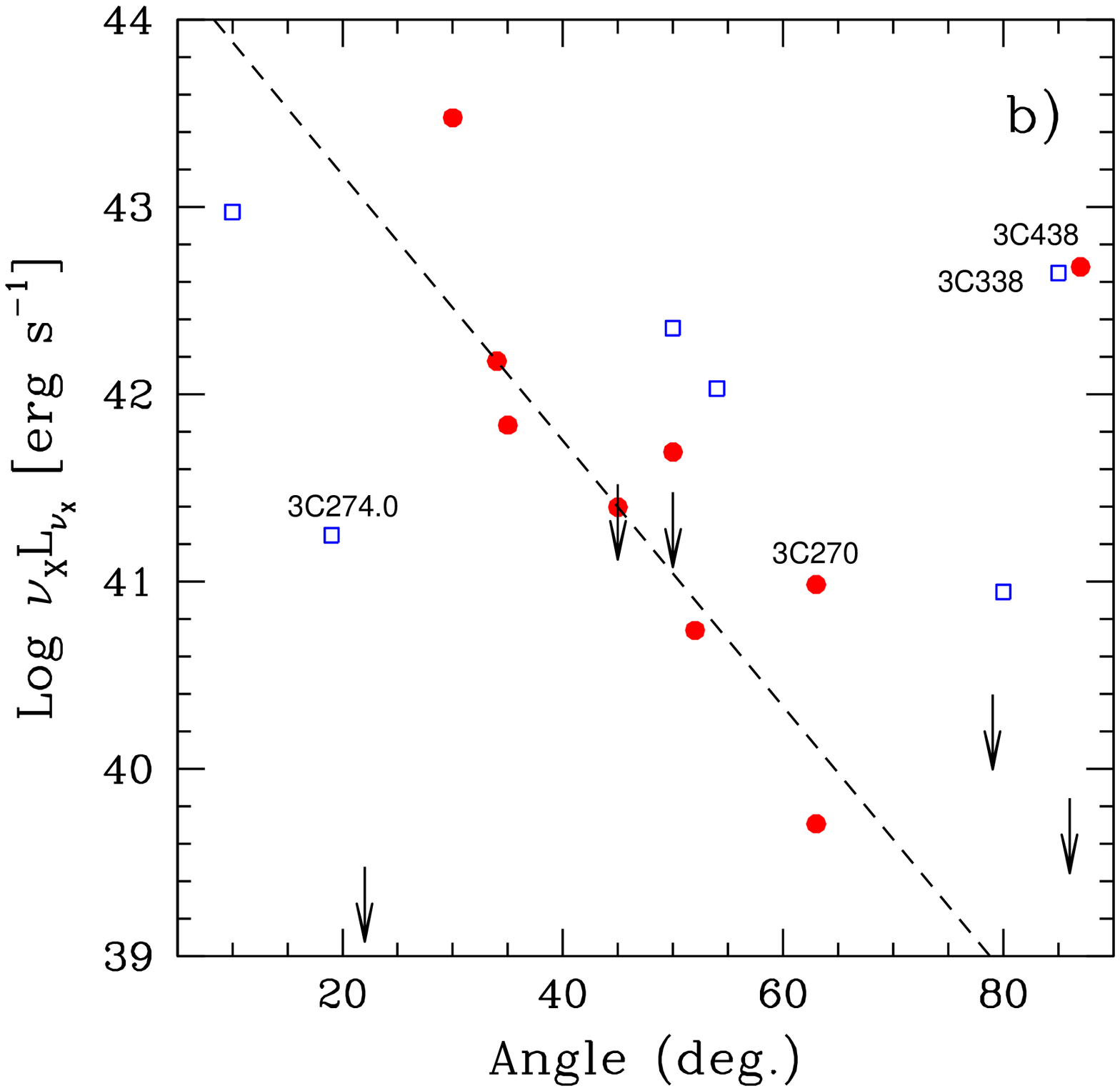}
\includegraphics[bb=40 175 550 675,width=8.0cm]{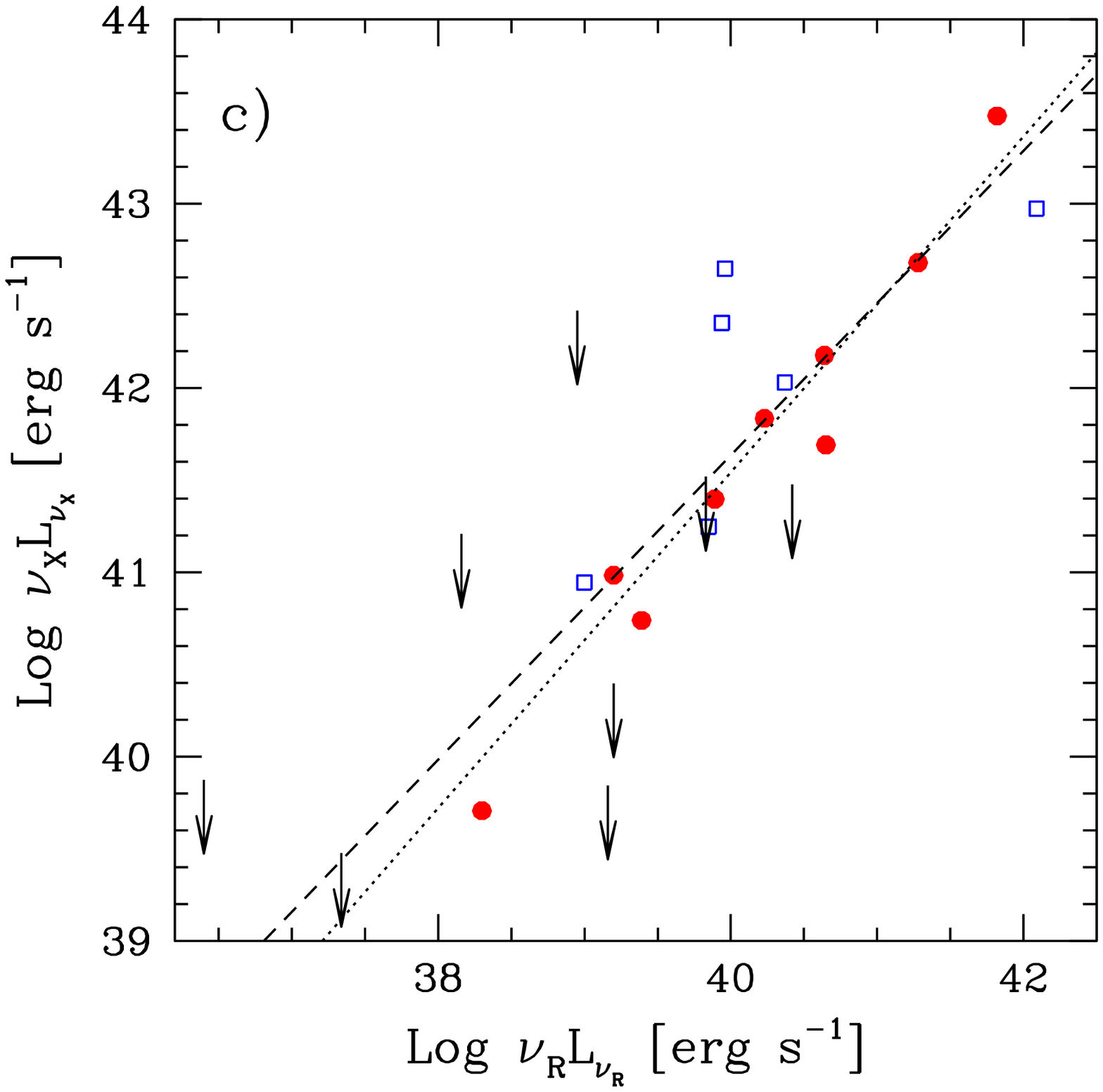}\includegraphics[bb=40 175 550 675,width=8.0cm]{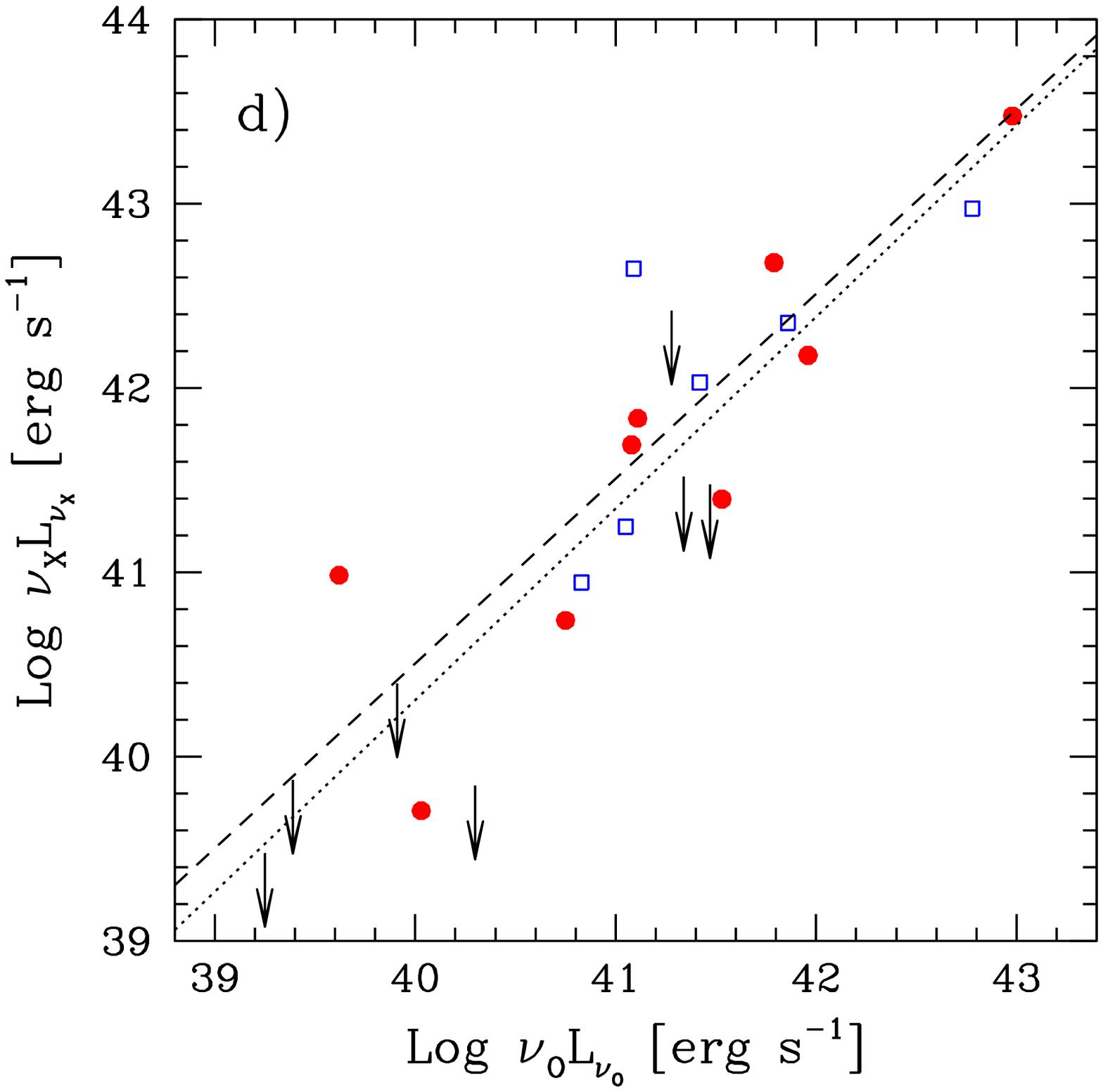}
\caption{Top panels: Plots of the inclination angle of the radio jet versus the
intrinsic column density from the X-ray spectral fits {\it (a)} and
versus the core X-ray luminosity in 0.3--8 keV {\it (b)}.  Bottom
panels: Plots of the core X-ray luminosity versus the radio luminosity
{\it (c)} and the optical luminosity {\it (d)}. 
Filled circles are
parameters from the \chandra\ data, open squares from the \xmm\
data. The dashed line is from a linear regression analysis of the firm
detections, while the dotted line is from an analysis including upper
limits.}
\label{correl}
\end{figure}

Figure~\ref{correl}a shows the plot of the intrinsic column density
N$_H^z$ (Table~6) versus the inclination angle of the jet
(Table~1). According to the unification schemes, the obscuring torus
becomes more prominent for larger viewing angles, so we would expect a
trend of larger N$_H^z$ for larger angles. No such trend is present in
Figure~\ref{correl}a, over 3 decades in N$_H^z$ and a factor 4 in
angle. 
We conclude that FRIs lack a standard
thick obscuring torus, in contrast with the expectations from the
unification models. 

In Figure~\ref{correl}b, we show the plot of the intrinsic X-ray
luminosity of the AGN versus the inclination angle of the radio
jet. Formally, a linear regression analysis shows that there is no
significant correlation. However, if the outliers sources 3C~338 and
3C~438 are excluded, a marginally significant correlation is detected, 
$P_{\rm r}(N)$ =2.3\% The correlation becomes statistically 
significant, $P_{\rm r}(N)$ =0.5\%, if only
the \chandra\ data (without 3C~438) are considered. The trend is in
the sense of decreasing X-ray luminosity with increasing angle, as
expected if a fraction of the X-ray emission is 
anisotropic, for example related to a
beamed component. Indeed, the X-ray luminosity also correlates tightly
with the radio and optical luminosities (Figures~\ref{correl}c and
\ref{correl}d), for which an origin from the unresolved jet was argued
(CH99; Hardcastle \& Worrall 2000; CA02)

As for the outliers, we note that 3C~438 are classified as an FRI
radio galaxy by CH99 but as an FRII by \citet{Rawl}: 
it is thus possible that this source is an intermediate FRI/II. 
In the case of 3C~338, the nucleus shows a high X-ray to submillimiter 
luminosity ratio compared to other 3C radio sources \citep{Quil}.
Both sources can thus be considered ``anomalous'' in the present sample.

A label in Figure~\ref{correl}b marks the position of 3C~274.0 (M87) and
3C~270. The first source is interesting because it was argued recently
on the basis of the \chandra\ data that the X-ray core flux originates
from the base of the jet \citep{Mars}; however, in the
Figure~\ref{correl}b this source has a deficit of X-ray emission. 
Variability could
possibly account for this discrepancy \citep{Harr}. 
On the other hands, 3C~270 appears
to have an excess of X-ray flux for its given inclination, supporting
our previous claims that the bulk of the X-ray emission
originates from the accretion flow \citep{Glio03}.

It is worth noting that several issues can affect the angle-to-$L_{\rm X}$ 
correlation, as well as the angle-to-N$_H^z$ correlation: 
a) uncertainties on the angles, which are rarely reported in the 
literature or are poorly determined.
This is shown for example by a detailed study of 
VLBI observations of a complete sample of radio galaxies 
from the B2 and 3CR catalogs \citet{Giov}, where the
errors range from a few to dozens of degrees; b) beaming effects, which 
should not play an important role at relatively large angles;
c) the possible concentration of obscuring material on the
equatorial plane, which can cause an intrinsically isotropic
emission to appear anisotropic.
For these reasons the origin of the X-ray radiation, 
that is, the fraction of X-rays produced by the jet and by the
accretion process, remains an open question.

\section{Discussion}

\subsection{Obscuration in FRIs}

An intriguing result of this paper is the finding that FRIs have
little or no excess X-ray absorption in their cores. This result fits
into the current debate, sparked by recent optical results, of whether
or not low-power radio galaxies have an obscuring pc-scale torus. As
mentioned earlier, \hst\ images showed the presence of a compact core
in the majority of 3C and B2 FRIs (CH99, CA02). However, based on the
same data, it was argued by \citet{Cao} that a standard torus is
indeed present in FRIs, obscuring most of the isotropic radiation from
the nucleus, and that the detected unresolved optical core is the
emission of the jet on scales of tens of parsec. The \chandra\ and
\xmm\ data support CH99 conclusions that no pc-scale 
torus is present in FRIs.

We now examine the possibility that the {\it direct} X-ray emission
from the core is blocked by a Compton-thick torus ($N_H^z > 10^{24}$
\nh), and that the measured X-ray radiation is due to reflection
toward the observer by a ``mirror'' located above the torus, as
postulated for Seyfert 2 galaxies (e.g., Matt et al. 2000).
In this case, the values of $N_H^z$ in Table~\ref{spec} would not
be associated to the torus, and an alternative explanation would have
to be found.

To test the Compton-thick torus hypothesis, we calculated for each
source of our sample the ratio $T=L_{X}/L_{OIII}$, where $L_{X}$ is
the unabsorbed 2-10 keV luminosity of the CCCX and $L_{OIII}$ is the
dereddened \ion{O}{3} luminosity from the literature. The latter is
produced in the Narrow-Line Region and is considered a good indicator
of the intrinsic AGN power. The average value for our FRI sample is
$T_{\rm FRI}=34.4 \pm 8.0$, where the uncertainty is the standard
dispersion. The average value of $T$ for the FRIs of our sample was
compared to the value derived for a sample of 9 Seyfert 1 galaxies and
for a sample of 16 Compton-thick Seyfert 2 galaxies. The data were
derived from \citet{Nand}, \citet{Bass}, and \citet{Ho}. We find
$T_{\rm Sy1}=37.8\pm9.6$ and $T_{\rm Sy2}=3.7\pm1.1$. The average $T$
for the FRIs of our sample is much larger than in Compton-thick
Seyfert 2s, and actually consistent with Seyfert 1s. This result
supports the idea that FRI radio galaxy lack a Compton-thick torus.

The detection of the compact X-ray core is not correlated with the
presence of a dust lane. In fact, dust lanes are equally present in
sources with ($\sim$55\%) and without ($\sim$45\%) a CCCX in our
sample \citep{Deko,Deru}. In the cases where excess absorption is
detected in the X-ray spectra and a dust lane is present, the column
density from the X-rays and that due to the gas in the lane,
N$_{H,V}$, are different. More precisely, comparing the values of
N$_{H,V}$ from \citet{Deko} and \citet{Deru} to
the values of N$_H^z$ in Table~\ref{spec}, we find that the former is
always about one order of magnitude smaller than the latter. This
discrepancy was already reported previously for 3C~270 (e.g.,
Chiaberge et al. 2003) and is known to exist for other AGN
\citep{Maia}. There are two possible interpretations: either the ratio
of gas-to-dust in AGN is different than in the Galaxy \citep{Maia}, or
the optical and X-ray extinction occur in distinct media
\citep{joe}. The latter hypothesis was suggested by us for 3C~270
\citep{Samb03}.

Thus, we conclude that FRIs lack a standard molecular torus,
confirming the optical results. As discussed by CH99 and other
authors, this implies that the lack of broad optical lines in these
sources can not be due to obscuration, but to the absence of ionizing
UV radiation, indicating inefficient accretion.  Alternatively, the
torus in FRIs may have a much smaller opening angle than in FRIIs,
and/or lack significant cold gas. A survey of low-power radio galaxies
in the far IR can also probe independently the presence of obscuration
in the cores of these sources.

\subsection{Origin of the X-rays in FRIs }

The origin of the X-rays in radio galaxies in general, and in FRIs in
particular, is still matter of considerable debate.  The strong
correlation between radio and X-ray core luminosities observed in
low-luminosity (e.g., Fabbiano et al. 1984; Canosa et al. 1999) and
high-luminosity (e.g., Worrall et al. 1994; Hardcastle et al. 1998)
radio galaxies has often been used to argue in favor of a common
origin from the unresolved base of the jet.  However, such a
correlation does not necessarily imply a common origin for the
radiation in the X-ray and radio regimes. Indeed, accretion processes
and relativistic jets are widely believed to be correlated phenomena
(e.g. Begelman et al. 1984) and, thus, a correlation between jet and
accretion-related fluxes is naturally expected.  In fact, \citet{Merl}
have recently demonstrated that the correlation between X-ray and
radio fluxes derives from a more general correlation (although with a
substantial scatter) involving also the black hole masses (the
so-called ``fundamental plane''), where the radio is produced by the
jet, whereas the X-rays are likely to be related to a radiatively
inefficient accretion flow. 

Previous studies of low-power AGN with \chandra\ demonstrate the importance 
of X-rays to make progress in this field (e.g., Di Matteo et al. 2001, 2003; 
Pellegrini et al. 2003; Terashima \& Wilson 2003;).
However, despite the high quality data provided by \chandra, the nature of 
the central engine in low-power AGN is still poorly known. Some
authors (e.g., Fabbiano et al. 2003; Pellegrini et al. 2003) favor a
jet-dominated scenario, where the entire spectral energy distribution (SED)
can be  explained in terms of non-thermal emission from the unresolved
base of a jet. Others (e.g., Di Matteo et al. 2003; Ptak et al. 2004)
favor a radiatively inefficient scenario where two solutions are possible:
a) a high accretion rate $\dot M$ (of the order of $\dot M_{\rm Bondi}$)
coupled with an extremely low radiative efficiency (basic ADAF scenario;
see, e.g., Narayan 2002); or b) a moderate  $\dot M$ 
($\ll \dot M_{\rm Bondi}$) combined with a moderately low efficiency
(more general RIAF scenario; see, e.g., Quataert 2003).

One of the reasons for this controversy is due to the poor
discriminating power of the low-power AGN spectral data, which does
not allow one to choose between the competing scenarios. In order to
break this spectral degeneracy, additional model-independent
constraints are required. Such information can be provided by the
X-ray temporal and spectral variability properties, as well as by
energetic considerations derived from the radio jet. This is the
approach adopted by \citep{Glio03} to investigate the origin of the
X-rays in the nuclear region of the nearby 3C~270. They found that the
bulk of the X-ray emission originates from a radiatively inefficient
accretion flow, with negligible jet contribution.

On the other hands, in this paper we find a correlation between the
X-ray core flux of FRIs and the inclination angle of the jet. The
trend is of decreasing X-ray luminosity with increasing orientation
angle. This is expected if a fraction of the X-rays is beamed. It is
thus possible that {\it both} the jet and a (radiatively inefficient;
see below) accretion flow contribute to the production of the X-rays,
with variable relative contributions. In support of this hypothesis,
we note that in Figure~\ref{correl}b 3C~270 has an excess of X-rays 
over what expected from the jet, while M87 has a strong deficit. 

More insightful results can be obtained from the spatial analysis, due
to the unprecedented spatial resolution of \chandra\ that allows to
disentangle the different components and to study the very inner
region of the AGN.  The first important results of this work is that a
substantial fraction (9 out of 25) of FRI radio galaxies does not show
a point-like component at high significance level (see Table~\ref{hardness}).  

The lack of CCCX among sources with an optical core may suggest a
different origin between X-ray and optical emission. The latter was
attributed to the base of the jet (CH99, Cao \& Rawlings 2004). 
Further support
to this hypothesis was lent by \citet{Khar} who demonstrated the
presence of a strong correlation between the optical point-like
emission and the radio core dominance. On the other hands, the X-rays
could be entirely produced in a very inefficient accretion flow. 

Alternatively, the X-rays originate from the base of the jet, as
suggested by the correlation we find between the inclination angle and
the X-ray flux (Figure~\ref{correl}b). 
In this case, the lack of CCCX in sources
with optical cores would place interesting constraints on the energy
of the relativistic electrons and/or on the jet magnetic field. 

\subsection{Accretion in FRIs}

The nature of the accretion process taking place onto the black hole
in the nuclei of FRIs is another unanswered question. Previous studies
of FRIs at X-ray (e.g., Sambruna et al. 1999; Di Matteo et al. 2001,
2003; Gliozzi, Sambruna, \& Brandt 2003) 
favor an inefficient accretion process on
the basis of the spectral energy distribution, energetic and temporal
properties.  However, for this specific sample of FRIs, \citet{Cao} on
the basis of \hst\ observations and model-dependent theoretical
considerations, have claimed that the accretion is efficient and takes
place in form of a standard accretion disk \citep{Shak}.  A first
model-independent way to test the nature of the accretion process at
work in the FRI nuclei is based on the comparison between the
Eddington and bolometric luminosities. In the following we assume the
X-rays are due to the accretion process. The limits become even more
severe if the X-rays are due in part or totally to the jet (see above). 

The Eddington luminosity is readily obtained once the black hole mass
is known.  Luckily, the black hole masses of all the objects of our
sample have been reported in literature (see Table~\ref{sources}).
The bolometric luminosity has been estimated from the X-ray luminosity
of the CCCX (i.e., the X-ray luminosity of the power-law component)
assuming the canonical bolometric correction of 10 (e.g., Elvis et
al. 1994). Incidentally, we note that the bolometric luminosities
derived in this way are fully consistent with the ones derived from
optical luminosities by \citet{Marc}. The values of $L_{\rm
bol}/L_{\rm Edd}$ for our sample, reported in Table~\ref{accretion},
are quite low, ranging between 2$\times 10^{-3}$ and 3$\times
10^{-8}$. These values are fully consistent with radiatively
inefficient scenario and clearly inconsistent with a ``standard''
accretion disk, as in the case of Seyfert 1s, where the bolometric
luminosity is a significant fraction (typically 10-30\%) of the
Eddington value. 
  
An alternative model-independent method to assess the efficiency of
the accretion process in FRIs is based on the direct estimate of the
radiative efficiency $\eta$.  This quantity can be readily obtained
from the formula $L_{\rm bol} = \eta \dot M_{\rm accr}c^{2}$, where a
rough estimate of the accretion rate is given by the Bondi accretion
rate $\dot M_{\rm Bondi} = 4\pi R^{2}_{\rm A}\rho_{\rm A}c_{\rm s}$,
with accretion radius $R_{\rm A}\sim GM/c_{\rm s}^{2}$, $c_{\rm s}$
the sound speed, and $\rho_{\rm A}$ the density at the accretion
radius.

\begin{figure}
\centering
\includegraphics[width=8.0cm]{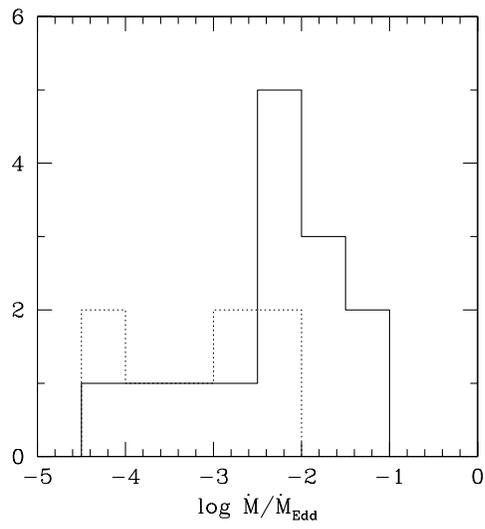}
\caption{Histogram of the accretion rate in Eddington units. The solid 
line refers to sources with confirmed CCCX; the dotted line refers to
sources with upper limit to CCCX. The Eddington accretion rate has been
inferred assuming a canonical radiative efficiency of 0.1.}
\label{istomdot}
\end{figure}

The value of $\rho_{\rm A}$ can be derived either from a spatial
method based on the deprojection of the surface brightness profile, or
from spectral method which makes use of the normalization of the
thermal spectral component (see, e.g., Gliozzi et al. 2003, 2004 for a
detailed description of either method).  For \chandra\ observations
with good quality brightness profiles, extending close to $R_{\rm A}$
the spatial method gives the most reliable results. On the other hand,
for \xmm\ observations the spectral method is preferable.  The values
of $\dot M_{\rm Bondi}$ and $\eta$ are reported in
Table~\ref{accretion}, and provide further support to the hypothesis
that the accretion in FRIs is radiatively inefficient if X-rays come
from accretion flow.  To visualize the results we plot the histogram
of the accretion rate in Eddington units in Figure~\ref{istomdot}. The
Eddington accretion rate has been inferred assuming a canonical
radiative efficiency of 0.1.

The only source with $\eta$ marginally consistent with the standard
accretion rates is 3C~28. However, in this case, as for all the
sources in Table~\ref{accretion}b, the values reported must be
considered carefully. For this source there is not a CCCX detection
and an upper limit of the X-ray luminosity has been used in the
evaluation of $L_{\rm bol}$ and $\eta$.

\section{Summary}

We have presented \chandra\ and \xmm\ observations of FRIs from the
3CR and B2 catalogs, for which there are high-quality \hst\ observations.
The main findings of this paper can be summarized as follows:

$\bullet$ A thorough spatial analysis reveals that 13 out of the 25 
objects in the sample exhibit a X-ray Central Compact Core
(CCCX), for 3 of the sources the detection of a CCCX is uncertain, 
and for the remaining 9 sources no CCCX is found. All
sources with a compact X-ray core also possess a compact core in the
optical; however, some sources with an optical core lack a CCCX. 

$\bullet$ All the FRIs are embedded in an extended component, which
is fit by at least one $\beta$-model with parameters 
typical for a host galaxy and/or for an intracluster medium. 

$\bullet$ The results from the spectral analysis are in good 
agreement with the spatial analysis. All the CCCX spectra are
well fit by power laws with photon indices 
$\Gamma \sim 1.1 - 2.6$, and at least one thermal component 
at softer energies. The remaining sources require only thermal 
components. 

$\bullet$ Among the sources with CCCX, intrinsic absorption 
over the Galactic value is required in only 8 cases, with 
N$_H^z \sim 10^{20} - 10^{21}$ \nh. 
This result, combined with model-independent tests, 
supports the previous claim (e.g. Chiaberge, Capetti, \& Celotti 1999) 
that low-power
radio galaxies lack a standard molecular torus on pc-scales. 
As a consequence, the non-detection of CCCX cannot be
attributed to obscuration effects, but rather to the 
intrinsic weakness of the AGN component.

$\bullet$ The origin of the X-rays (i.e., jet versus accretion)
is still an open question. On the one hand, the non-detection of 
CCCX in objects with an optical CCC (interpreted as jet radiation)
may indicate a different origin between X-ray and optical emission.
On the other hand, the correlation between the \chandra\ X-ray 
luminosity and the inclination
angle of the jet suggests that at least a fraction of the X-ray
emission originates in the jet. However, the accretion flow can 
still play a role in the production of the X-rays in individual 
sources, as shown by our previous analysis of 3C~270. 

$\bullet$ If the detected X-ray luminosities are considered as an
upper limit on the isotropic component, 
stringent limits on the Eddington ratios
$L_{bol}/L_{Edd}$ and on the accretion efficiency $\eta$ can be derived,
with $L_{bol}/L_{Edd} \sim 10^{-3}-10^{-8}$ and $\eta \sim
10^{-2}-10^{-6}$, suggesting radiatively inefficient accretion. The
upper limits are even lower if part or all the X-rays are due to
the jet.

\begin{acknowledgements}

We gratefully acknowledge the financial support provided by NASA grant
G04-5115A (DD), and NASA LTSA grant NAG5--10708 (RMS, MG). RMS gratefully
acknowledges support from an NSF CAREER award and from the Clare
Boothe Luce Program of the Henry Luce Foundation.

\end{acknowledgements}

\clearpage

\begin{table} 
\caption{The Sample} 
\begin{center}
\begin{tabular}{l c c c c c c c c }  
\hline
\hline        
Source    &  Other name  & $z$ & $N_{\rm H,Gal}$& Angle & Ref. & $Log M_{\rm BH}$ &  $Log (\nu_{r}L_{\nu_{r}})$    &  $Log (\nu_{o}L_{\nu_{o}})$   \\
~~(1)  &      (2)      & (3) &    (4)         &  (5)  & (6)  &   (7)        &  (8)       &  (9)         \\   
\hline       
3C~28       &           &   0.1953   &  5.40  &          &        & 8.67       &  38.95   &   41.28  \\ 
B2~0055+30  & NGC~315   &   0.0165   &  5.86  &  35$^{a}$& Gi01   & 9.18       &  40.23   &   41.11  \\ 
3C~31       & NGC~383   &   0.0170   &  5.36  &  52      & Ha02   & 7.89       &  39.39   &   40.75  \\ 
B2~0120+33  & NGC~507   &   0.0165   &  5.25  &          &        & 8.94       &  36.40   &   39.39  \\ 
B2~0149+35  & NGC~708   &   0.0162   &  5.29  &          &        & 8.46$^{b}$ &  38.16   &   Dusty  \\ 
3C~66B      &           &   0.0213   &  9.15  &  45      & Ha01   & 8.84       &  39.89   &   41.53  \\ 
3C~78       & NGC~1218  &   0.0287   & 10.70  &  30      & Sp95   & 8.98       &  40.87   &   42.43  \\ 
3C~84       & NGC~1275  &   0.0176   & 15.70  &  10      & Pe90   & 9.28       &  42.09   &   42.78  \\ 
B2~0755+37  & NGC~2484  &   0.0428   &  5.07  &  34      & La99   & 8.93       &  40.64   &   41.96  \\ 
3C~264      & NGC~3862  &   0.0217   &  2.19  &  50      & Ba97   & 8.85       &  39.94   &   41.86  \\ 
3C~270      & NGC~4261  &   0.0075   &  1.52  &  63      & Pi01   & 8.57       &  39.20   &   39.62  \\ 
3C~272.1    & M~84      &   0.0035   &  2.63  &  63$^{a}$& Gi01   & 8.35       &  38.30   &   40.03  \\ 
3C~274.0    & M~87      &   0.0044   &  2.59  &  19      & Gi01   & 8.26       &  39.85   &   41.05  \\ 
B2~1256+28  & NGC~4869  &   0.0229   &  0.89  &          &        & 8.12$^{b}$ &  37.69   &   Dusty  \\ 
B2~1257+28  & NGC~4874  &   0.0241   &  0.89  &  22$^{a}$& Fe87   & 8.63       &  37.34   &   39.25  \\ 
B2~1346+26  & 4C26.42   &   0.0633   &  1.18  &  45      & Ge93   & 9.02       &  39.83   &   41.34  \\ 
3C~317      & UGC~9799  &   0.0345   &  2.90  &  50$^{a}$& Ve00   & 8.80       &  40.65   &   41.08  \\ 
3C~338      & NGC~6166  &   0.0304   &  0.84  &  85      & Gi01   & 9.23       &  39.96   &   41.09  \\ 
3C~346      &           &   0.1620   &  5.67  &  30      & Gi01   & 8.89       &  41.82   &   42.98  \\ 
3C~348      & Her A     &   0.1540   &  6.28  &  50$^{a}$& Sa02   & 8.84       &  40.42   &   41.47  \\ 
B2~2116+26  & NGC~7052  &   0.0156   & 12.90  &  79      & La99   & 8.60       &  39.20   &   39.91  \\ 
3C~438      &           &   0.2900   & 19.70  &  87$^{a}$& Be95   & 8.80       &  41.28   &   41.79  \\ 
3C~449      & UGC~12064 &   0.0171   &  12.00 &  80      & Ha98   & 8.42$^{c}$ &  39.00   &   40.83  \\ 
B2~2236+35  & UGC~12127 &   0.0276   &  10.00 &  86      & La99   & 8.55       &  39.16   &   40.30  \\ 
3C~465      & NGC~7720  &   0.0302   &   4.84 &  54      & Gi01   & 9.32       &  40.37   &   41.42  \\ 
\hline

\end{tabular}
\end{center}
{\bf Column Explanation}: 
1=Source name; 2=Alternative name; 3=Redshift; 4=Galactic 
column density in $10^{20}$ \nh, from Dickey \& Lockman (1990); 
5=Inclination angle in degrees of the jet axis with respect the line of 
sight ($^{a}$=Average value); 6=Reference for column 5. Ba97=Baum 1997; 
Be95=Best 1995; Fe87=Feretti 1987; 
Ge93=Ge 1993; Gi01=Giovannini 2001; Ha98=Hardcastle 1998; Ha01=Hardcastle 2001;
Ha02=Hardcastle 2002; La99=Laing 1999; Pe90=Pedlar 1990; Pi01=Piner 2001; 
Sa02=Saxton 2002; Sp95=Sparks 1995; Ve00=Venturi 2000.
7=Black hole mass in solar masses (Marchesini et al. 2004; $^{b}$=Woo \& Urry 2003;
$^{c}$=Bettoni et al. 2003);
8=Core radio luminosity at 5 GHz, in \lum, from \citet{Coll}; 
9=Core optical luminosity, in \lum, from CH99, CA02. 

\label{sources}
\end{table}       

%\clearpage
\begin{table} 
\caption{Observation Log}
\begin{center}
\begin{tabular}{l c c c c } 
\hline
\hline        

Source      & Satellite        &   Date        & Net Exp. &  Count Rate  \\
~~(1)       &   (2)            &     (3)       &  (4)     &   (5)  \\
\hline        
3C~28       & {\it Chandra}   & 2002-10-07   &  48.73  &  0.023$\pm$0.001 \\
B2~0055+30  & {\it Chandra}   & 2003-02-22   &  51.28  &  0.108$\pm$0.001 \\ 
3C~31       & {\it Chandra}   & 2000-11-06   &  42.52  &  0.039$\pm$0.001 \\ 
B2~0120+33  & {\it Chandra}   & 2000-10-11   &  18.26  &  0.035$\pm$0.001 \\
            & {\it XMM}       & 2001-01-15   &  26.56  &  0.082$\pm$0.002 \\ 
B2~0149+35  & {\it Chandra}   & 2001-08-03   &  28.74  &  0.037$\pm$0.001 \\
            & {\it XMM}       & 2001-01-16   &  17.42  &  1.020$\pm$0.008 \\ 
3C~66B      & {\it Chandra}   & 2000-11-20   &  28.36  &  0.043$\pm$0.001 \\ 
3C~78       & {\it Chandra}   & 2001-12-16   &   5.23  &  0.224$\pm$0.007 \\ 
3C~84       & {\it XMM}       & 2001-01-30   &  24.71  & 11.648$\pm$0.022 \\ 
B2~0755+37  & {\it Chandra}   & 2000-04-03   &   5.61  &  0.098$\pm$0.004 \\ 
3C~264      & {\it XMM}       & 2001-05-26   &  24.12  &  0.228$\pm$0.003 \\ 
3C~270      & {\it Chandra}   & 2000-05-06   &  32.51  &  0.126$\pm$0.002 \\ 
3C~272.1    & {\it Chandra}   & 2000-05-19   &  28.19  &  0.075$\pm$0.002 \\ 
3C~274.0    & {\it Chandra}   & 2000-07-30   &  12.69  &  0.521$\pm$0.006 \\
            & {\it XMM}       & 2000-06-19   &  25.38  &  1.913$\pm$0.010 \\ 
B2~1256+28  & {\it XMM}       & 2001-12-04   &  22.61  &  $<$0.617$^{\mathrm{a}}$\\ 
B2~1257+28  & {\it Chandra}   & 1999-11-04   &   9.53  &  0.020$\pm$0.002 \\ 
B2~1346+26  & {\it Chandra}   & 2004-01-14   &  14.30  &  0.113$\pm$0.003 \\
3C~317      & {\it Chandra}   & 2000-09-03   &  36.20  &  0.077$\pm$0.001 \\ 
3C~338      & {\it Chandra}   & 1999-12-11   &  16.94  &  0.118$\pm$0.003 \\
            & {\it XMM}       & 2002-08-15   &   5.77  &  2.999$\pm$0.023 \\ 
3C~346      & {\it Chandra}   & 2002-08-03   &  39.75  &  0.079$\pm$0.001 \\ 
3C~348      & {\it Chandra}   & 2001-07-25   &  14.58  &  0.047$\pm$0.002 \\ 
B2~2116+26  & {\it Chandra}   & 2002-09-21   &   9.63  &  0.028$\pm$0.002 \\
3C~438      & {\it Chandra}   & 2002-12-27   &  37.06  &  0.016$\pm$0.001 \\
3C~449      & {\it XMM}       & 2001-12-09   &  16.71  &  0.087$\pm$0.002 \\ 
B2~2236+35  & {\it Chandra}   & 2001-05-13   &   9.39  &  0.021$\pm$0.001 \\ 
3C~465      & {\it XMM}       & 2002-06-22   &   4.48  &  0.148$\pm$0.006 \\ 
\hline
                      
\end{tabular}
\end{center}
$^{\mathrm{a}}$ 3$\sigma$ upper limit. 

{\bf Columns explanation}: 1=Source name; 2=Instrument; 3=Observation
date (yyyy-mm-dd); 4=Live time in ksec after data screening. For \xmm\
observations, the exposures refer to data taken with the EPIC pn
camera; 5=Total source count rate in the 0.3-8 keV energy range. The
extraction radius is 5\arcsec\ for \chandra\ and 30\arcsec\ for \xmm,
except for B2~0120+33 and 3C~264 (12\arcsec) and for 3C274.0
(9\arcsec).

\label{log}
\end{table}

%\clearpage
\begin{table} 
\caption{Results of the X-ray spatial analysis}
\begin{center}
\begin{tabular}{l c c c c c c} 
\hline        
\hline        
Source      & \multicolumn{2}{c}{CCC} & Core Sig. & Sat.&  Count Rate  \\
            &     opt.    &     X     &           &     &              \\
~~(1)       &     (2)     &    (3)    &  (4)      & (5) &   (6)        \\
\hline       			        	       		   
3C~28       & Up  &  N  & ~~~92.9\%  & C &         $<$0.001        \\
B2~0055+30  & Up  &  Y  & $>$99.9\%  & C &  0.035$\pm$0.001        \\
3C~31       & Y   &  Y  & $>$99.9\%  & C &  0.003$\pm$0.001        \\
B2~0120+33  & Up  &  N  & ~~~12.4\%  & C &         $<$0.001        \\
B2~0149+35  & D   &  N  & ~~~38.7\%  & C &         $<$0.001        \\
3C~66B      & Y   &  Y  & $>$99.9\%  & C &  0.006$\pm$0.001        \\
3C~78       & Y   &  U  &     ...    & C &  0.053$\pm$0.003        \\
3C~84       & Y   &  Y  & $>$99.9\%  & X &  2.815$\pm$0.011        \\
B2~0755+37  & Y   &  Y  & $>$99.9\%  & C &  0.009$\pm$0.001        \\
3C~264      & Y   &  U  &    ...     & X &  0.020$\pm$0.001        \\
3C~270      & Y   &  Y  & ~~~99.9\%  & C &  0.016$\pm$0.001        \\
3C~272.1    & Y   &  U  &    ...     & C &  0.005$\pm$0.001        \\
3C~274.0    & Y   &  Y  & $>$99.9\%  & C &  0.052$\pm$0.002        \\
B2~1256+28  & D   &  N  &    ...     & X & $<$0.160$^{\mathrm{a}}$ \\ 
B2~1257+28  & Up  &  N  & ~~~30.2\%  & C &         $<$0.001        \\
B2~1346+26  & Y   &  N  & ~~~22.3\%  & C &  0.002$\pm$0.001        \\ 
3C~317      & Y   &  Y  & $>$99.9\%  & C &  0.006$\pm$0.001        \\
3C~338      & Y   &  Y  & ~~~99.9\%  & C &  0.002$\pm$0.001        \\
3C~346      & Y   &  Y  & $>$99.9\%  & C &  0.017$\pm$0.001        \\
3C~348      & Y   &  N  & ~~~21.4\%  & C &         $<$0.001        \\
B2~2116+26  & Y   &  N  & ~~~95.6\%  & C &  0.002$\pm$0.001        \\
3C~438      & Up  &  Y  & $>$99.9\%  & C &         $<$0.001        \\
3C~449      & Y   &  Y  & $>$99.9\%  & X &  0.012$\pm$0.001        \\
B2~2236+35  & Y   &  N  & ~~~15.4\%  & C &         $<$0.001        \\
3C~465      & Y   &  Y  & $>$99.9\%  & X &  0.023$\pm$0.002        \\
\hline
                      
\end{tabular}
\end{center} 
$^{\mathrm{a}}$ 3$\sigma$ upper limit. 

{\bf Columns Explanation}: 1=Source name; 
2=Detection of the optical Central Compact Core: 
Y=Yes, D=Dusty galaxy; Up=Upper limit (CH99, CA02); 
3=Detection of the X-ray Central Compact Core: Y=Yes, N=No,
U=Uncertain;
4=Significance of the PSF, from the F-test; 
5=C: \chandra; X: \xmm;  
6=X-ray count rate of the core in the energy 
range 2--10 keV from an extraction radius of 1.5\arcsec. 

\label{hardness}
\end{table}       

%\clearpage
\begin{table} 
\caption{Results of fits of radial profiles}
\begin{center}
\begin{tabular}{l c c c } 
\hline        
\hline        
Source      &    Core Radius   &     $\beta$     &         Norm.      \\
~~(1)       &         (2)      &      (3)        &          (4)       \\
\hline       			        	       		   
3C~28                     &   6.24$\pm$0.58  &  0.42$\pm$0.04  &    3.2$\pm$0.1   \\
B2~0055+30                &   1.98$\pm$0.20  &  0.54$\pm$0.03  &   23.8$\pm$3.2   \\
3C~31                     &   3.46$\pm$0.39  &  0.68$\pm$0.10  &    7.0$\pm$0.8   \\
B2~0120+33$^{\mathrm{a}}$ &   0.59$\pm$0.09  &  0.46$\pm$0.04  &   52.3$\pm$7.5   \\
B2~0149+35                &  11.16$\pm$0.54  &  0.42$\pm$0.02  &    5.0$\pm$0.1   \\
3C~66B                    &   0.49$\pm$1.02  &  0.49$\pm$0.07  &   45.4$\pm$17.0  \\
3C~84                     & 104.09$\pm$0.76  &  0.64$\pm$0.01  &   11.6$\pm$0.1   \\
B2~0755+37                &   0.59$\pm$0.59  &  0.60$\pm$0.16  &  107.5$\pm$107.5 \\
3C~270                    &   0.98$\pm$0.07  &  0.52$\pm$0.01  &  126.7$\pm$14.5  \\
3C~274.0                  &   0.22$\pm$0.20  &  0.24$\pm$0.01  &  104.6$\pm$43.2  \\
B2~1257+28                &   1.48$\pm$0.15  &  0.67$\pm$0.01  &   16.7$\pm$2.9   \\
B2~1346+26                &   3.32$\pm$0.54  &  0.25$\pm$0.01  &   16.1$\pm$0.9   \\ 
3C~317                    &  38.61$\pm$0.22  &  0.67$\pm$0.01  &    5.8$\pm$0.1   \\
3C~338                    &  12.91$\pm$0.36  &  0.34$\pm$0.01  &   11.5$\pm$0.2   \\
3C~346                    &   0.10$\pm$0.10  &  0.30$\pm$0.06  &    7.3$\pm$6.6   \\
3C~348$^{\mathrm{a}}$     &   4.05$\pm$0.70  &  0.75$\pm$0.30  &   11.5$\pm$1.1   \\
B2~2116+26                &   1.11$\pm$0.28  &  0.48$\pm$0.03  &   17.5$\pm$6.2   \\
3C~438$^{\mathrm{a}}$     &  13.81$\pm$2.89  &  0.87$\pm$0.61  &    1.9$\pm$0.1   \\
3C~449                    & 106.20$\pm$54.62 &  0.71$\pm$0.14  &        $<$0.1    \\
B2~2236+35                &   0.74$\pm$0.15  &  0.48$\pm$0.04  &   26.7$\pm$6.4   \\
3C~465                    & 194.50$\pm$43.44 &  0.60$\pm$0.34  &         $<$0.1   \\
\hline
                      
\end{tabular}
\end{center}

$^{\mathrm{a}}$ An additional $\beta$-model is necessary.

{\bf Columns Explanation}: 1=Source name; 2=Core radius in arcseconds; 3=$\beta$ 
parameter; 4=Normalization of the $\beta$-model in units of $10^{-4}$ 
counts s$^{-1}$ arcsec$^{-2}$. 
Uncertain detections (3C~78, 3C~264, and 3C~272.1) are not considered 
\label{beta}
\end{table}       

%\clearpage
\begin{table} 
\caption{Core X-ray spectral analysis} 
\begin{center}
\begin{tabular}{l c | c c c c | c c | c} 
\hline
\hline        
&&\multicolumn{4}{c}{Thermal Model}&\multicolumn{2}{c}{Power-law}&\\
\hline        

Source       &  &         kT$_1$               & $Z_1$ &  kT$_2$               
& $Z_2$ &$N_H^z$      & $\Gamma$     &  $\chi^2_{\rm red}/$d.o.f.\\
~~(1)  & (2) &  (3)  &  (4) &  (5) &  (6)  &  (7) & (8) &  (9) \\
\hline            

\hline
\multicolumn{9}{c}{a) Confirmed CCCX}\\
\hline

B2~0055+30   & C &  0.51$^{+0.05}_{-0.05}$  & 1.0 & ...                    & ... &  72.7$^{+18.4}_{-10.5}$  &  1.56$^{+0.17}_{-0.09}$  & 0.97/153   \\
3C~31        & C &  0.69$^{+0.06}_{-0.05}$  & 0.2 & ...                    & ... &  ...                     &  1.22$^{+0.27}_{-0.19}$  & 1.03/33   \\
3C~66B       & C &  0.36$^{+0.16}_{-0.07}$  & 1.0 & ...                    & ... &  ...                     &  2.17$^{+0.14}_{-0.15}$  & 0.91/45   \\
3C~84$^{\mathrm{a}}$& X &  2.38$^{+0.37}_{-0.40}$  & 1.0 & 0.82$^{+0.17}_{-0.47}$ & 0.2 &   8.0$^{+ 2. }_{- 1.6}$  &  1.86$^{+0.05}_{-0.05}$  & 1.29/460 \\
B2~0755+37   & C &  0.26$^{+0.14}_{-0.18}$  & 0.2 & ...                    & ... &   ...                    &  2.18$^{+0.28}_{-0.19}$  & 0.67/15   \\
3C~270       & C &  0.60$^{+0.03}_{-0.03}$  & 1.0 & ...                    & ... & 702.6$^{+238.7}_{-207.2}$&  1.09$^{+0.44}_{-0.23}$  & 1.04/60   \\
3C~274.0$^{\mathrm{b}}$& X &  0.10$^{+0.94}_{-0.02}$  & 0.2 & 1.66$^{+0.05}_{-0.05}$ & 1.0 &   2.3$^{+ 1.5}_{- 1.5}$  &  2.40$^{+0.08}_{-0.06}$  & 1.31/575 \\
3C~317       & C &  0.18$^{+0.09}_{-0.04}$  & 1.0 & 0.80$^{+0.68}_{-0.21}$ & 1.0 &  ...                     &  1.81$^{+0.13}_{-0.10}$  & 0.98/52   \\
3C~338       & X &  2.23$^{+0.18}_{-0.16}$  & 1.0 & 0.15$^{+0.06}_{-0.04}$ & 1.0 &  15.4$^{+ 4.7}_{- 3.5}$  &  2.15$^{+0.16}_{-0.19}$  & 0.94/569  \\
3C~346       & C &  0.64$^{+0.40}_{-0.38}$  & 0.2 & ...                    & ... &  ...                     &  1.69$^{+0.09}_{-0.09}$  & 1.06/101  \\
3C~438       & C &  ...                     & ... & ...                    & ... &  ...                     &  1.54$^{+0.30}_{-0.26}$  & 231.1/321$^{\mathrm{c}}$ \\
3C~449       & X &  0.58$^{+0.15}_{-0.27}$  & 0.2 & 1.26$^{+0.26}_{-0.20}$ & 1.0 &  14.5$^{+117.9}_{-12.3}$ &  2.13$^{+0.65}_{-0.55}$  & 0.95/58   \\
3C~465       & X &  0.97$^{+0.08}_{-0.08}$  & 1.0 & ...                    & 1.0 &  25.7$^{+10.9}_{- 5.}$   &  2.59$^{+0.37}_{-0.20}$  & 0.86/41   \\

\hline
\multicolumn{9}{c}{b) Candidate CCCX}\\
\hline
3C~264       & X &  0.33$^{+0.08}_{-0.06}$  & 0.2 & ...                    & ... &   ...                    &  2.48$^{+0.04}_{-0.04}$  & 0.92/290  \\
3C~272.1     & C &  0.61$^{+0.14}_{-0.19}$  & 0.8 & ...                    & ... &  17.8$^{+69. }_{- 8. }$  &  2.06$^{+0.31}_{-0.29}$  & 0.61/28   \\
\hline
                      
\end{tabular}
\end{center}
$^{\mathrm{a}}$ Additional themal component necessary:
kT$_3$=0.17$\pm 0.02$,  $Z_3 \cong 0.6$

$^{\mathrm{b}}$ Additional themal component necessary: kT$_3$=0.81$\pm 0.02$,
$Z_3 \cong 0.5$ 

$^{\mathrm{c}}$ Values of C-statistics and PHA bins.

{\bf Columns Explanation}: 1=Source name; 2=Satellite data used for the
spectral analysis (C=\chandra, X=\xmm); 3-6=Temperature $kT$ in keV and 
abundance $Z/Z_{\odot}$; 
7-8=Absorption column density at the source's redshift, 
in $10^{20}$ cm$^{-2}$; 
Photon index $\Gamma$ of the power-law component; 
9=Reduced $\chi^2$ of the fit and degrees of freedom.
\label{spec}
\end{table}       

%\clearpage
\begin{table}
\caption{Sources X-ray Fluxes and Luminosities}
\begin{center}
\begin{tabular}{l  c c | c c} 
\hline
\hline        
&\multicolumn{2}{c}{Total}&\multicolumn{2}{c}{Power-law}\\
\hline        
Source      & Flux  & Lum. & Flux  & Lum. \\
~~(1)       &  (2)  &  (3) &  (4)  &  (5) \\
       
\hline       
\multicolumn{5}{c}{a) Confirmed CCCX}\\
\hline
B2~0055+30  &  0.88  &  0.72 &  0.84  &  0.69  \\ 
3C~31       &  0.14  &  0.08 &  0.10  &  0.05  \\ 
3C~66B      &  0.23  &  0.27 &  0.21  &  0.25  \\ 
3C~84       & 14.10  & 12.40 & 11.10  &  9.40   \\ 
B2~0755+37  &  0.39  &  1.73 &  0.35  &  1.51  \\ 
3C~264      &  2.24  &  2.35 &  2.12  &  2.26  \\ 
3C~270      &  0.59  &  0.11 &  0.49  &  0.10  \\ 
3C~272.1    &  0.16  &  0.01 &  0.14  &  0.01  \\ 
3C~274.0    &  7.28  &  0.32 &  3.76  &  0.18  \\ 
3C~317      &  0.22  &  0.60 &  0.20  &  0.49  \\ 
3C~338      &  4.45  &  9.57 &  1.66  &  4.44  \\ 
3C~346      &  0.48  & 31.05 &  0.47  & 30.24  \\ 
3C~438      &  0.04  &  4.78 &  0.04  &  4.78  \\
3C~449      &  0.17  &  0.16 &  0.09  &  0.09  \\ 
3C~465      &  0.32  &  1.18 &  0.25  &  1.07  \\ 
\hline
\multicolumn{5}{c}{b) Upper Limits to CCCX (3$\sigma$)}\\
\hline
3C~28       &  0.18  & 16.94 & $<$0.03  &  $<$2.64  \\
B2~0120+33  &  0.25  &  0.16 & $<$0.01  &  $<$0.01  \\ 
B2~0149+35  &  2.36  &  1.59 & $<$0.23  &  $<$0.16  \\ 
B2~1257+28  &  0.06  &  0.07 & $<$0.01  &  $<$0.01  \\ 
B2~1346+26  &  0.68  &  6.60 & $<$0.03  &  $<$0.33  \\ 
3C~348      &  0.09  &  5.91 & $<$0.01  &  $<$0.30  \\ 
B2~2116+26  &  0.07  &  0.04 & $<$0.05  &  $<$0.03  \\
B2~2236+35  &  0.03  &  0.14 & $<$0.01  &  $<$0.01  \\ 
\hline
                      
\end{tabular}
\end{center}

{\bf Columns Explanation}: 1=Source name; 2-3=Observed total flux and
intrinsic luminosity in 0.3--8 keV of the source; 
4-5=Observed flux and intrinsic luminosity in 0.3--8 keV for AGN. 
The fluxes are expressed in $10^{-12}$ \flux and the luminosities in 
$10^{42}$ \lum. 
\label{flux}
\end{table}       

%\clearpage
\begin{table}
\caption{Correlation probabilities and parameters}
\begin{center}
\begin{tabular}{c c c c c} 
\hline        
\hline        
 &  $P_{\rm r}$ & $|r|$ & a & b \\
& (1) & (2) & (3) \\
\hline  
\multicolumn{5}{c}{a) Detections only}\\
\hline
Angle / Log $N_{\rm H}$                      & 33.9  &  0.36 & 20.930 &  0.005 \\
Log $\nu_XL_{\nu_X}$ / Log $N_{\rm H}$       & 58.2  &  0.21 & 22.984 & -0.044 \\
Angle / Log $\nu_XL_{\nu_X}$                 & 42.4  &  0.22 & 42.937 & -0.025 \\
                                             &  0.5  &  0.87 & 44.590 & -0.071 \\
Log $\nu_RL_{\nu_R}$ / Log $\nu_XL_{\nu_X}$  & 6.8E-4&  0.89 &  8.596 &  0.826 \\
Log $\nu_OL_{\nu_O}$ / Log $\nu_XL_{\nu_X}$  & 6.2E-3&  0.85 &  0.427 &  1.002 \\
\hline						         
\multicolumn{5}{c}{a) Detections + Upper limits}\\
\hline
Angle / Log $\nu_XL_{\nu_X}$                 & 13.5  & ... & 42.519 & -0.018 \\
Log $\nu_RL_{\nu_R}$ / Log $\nu_XL_{\nu_X}$  &  0.0  & ... &  5.101 &  0.911 \\
Log $\nu_OL_{\nu_O}$ / Log $\nu_XL_{\nu_X}$  &  0.0  & ... & -1.253 &  1.039 \\
\hline
\end{tabular}
\end{center}

{\bf Columns Exaplanation}: 1=Probability (in \%) that the distribution is generated
by a random population. A small probability indicates a significant
correlation; 2=Linear coefficient from a linear regression analysis;
3-4=Coefficients of the linear regression parameters a and b (where y=a+bx).
\label{corr}
\end{table}       

%\clearpage
\begin{table} 
\caption{Accretion properties}
\begin{center}
\begin{tabular}{l c c c} 
\hline
\hline        
Source      & $\dot M_{\rm Bondi}$ & $L_{\rm bol}/L_{\rm Edd}$ &  $\eta$ \\
~~(1)       &      (2)        &       (3)       &  (4)    \\
       
\hline       
\multicolumn{4}{c}{a) Confirmed CCCX}\\
\hline
B2~0055+30                & 4.6 $10^{-1}$  & 2.6 $10^{-5}$  & 1.9 $10^{-4}$  \\ 
3C~31                     & 3.5 $10^{-4}$  & 5.3 $10^{-5}$  & 2.6 $10^{-2}$  \\ 
3C~66B                    & 3.9 $10^{-1}$  & 1.2 $10^{-5}$  & 4.5 $10^{-5}$  \\ 
3C~84$^{\mathrm{a}}$      & 3.0 $10^{-1}$  & 2.2 $10^{-4}$  & 3.1 $10^{-3}$  \\ 
B2~0755+37$^{\mathrm{a}}$ & 6.0 $10^{-1}$  & 5.9 $10^{-5}$  & 1.9 $10^{-4}$  \\ 
3C~264$^{\mathrm{a}}$     & 5.1 $10^{-2}$  & 6.7 $10^{-5}$  & 2.1 $10^{-3}$  \\ 
3C~270                    & 9.5 $10^{-2}$  & 2.1 $10^{-5}$  & 1.8 $10^{-4}$  \\ 
3C~272.1$^{\mathrm{a}}$   & 3.1 $10^{-1}$  & 8.3 $10^{-7}$  & 1.3 $10^{-6}$  \\ 
3C~274.0                  & 1.7 $10^{-2}$  & 2.2 $10^{-5}$  & 5.3 $10^{-4}$  \\ 
3C~317                    & 6.1 $10^{-2}$  & 3.6 $10^{-5}$  & 8.3 $10^{-4}$  \\ 
3C~338                    & 8.7 $10^{-3}$  & 8.7 $10^{-5}$  & 3.8 $10^{-2}$  \\ 
3C~346                    & 5.3 $10^{-2}$  & 2.0 $10^{-3}$  & 6.5 $10^{-2}$  \\ 
3C~449$^{\mathrm{a}}$     & 1.0 $10^{-4}$  & 5.7 $10^{-5}$  & 6.6 $10^{-2}$  \\ 
3C~465$^{\mathrm{a}}$     & 2.5 $10^{-2}$  & 9.4 $10^{-6}$  & 1.8 $10^{-2}$  \\ 
\hline
\multicolumn{4}{c}{b) Undetected CCCX}\\
\hline
3C~28       & 5.8 $10^{-4}$  & 1.2 $10^{-4}$  & 2.2 $10^{-1}$ \\
B2~0120+33  & 1.8 $10^{-1}$  & 5.0 $10^{-7}$  & 5.4 $10^{-6}$ \\ 
B2~0149+35  & 6.2 $10^{-4}$  & 9.0 $10^{-6}$  & 9.3 $10^{-3}$ \\ 
B2~1257+28  & 1.5 $10^{-2}$  & 3.2 $10^{-8}$  & 2.0 $10^{-6}$ \\ 
B2~1346+26  & 9.2 $10^{-3}$  & 1.4 $10^{-5}$  & 3.6 $10^{-3}$ \\ 
3C~348      & 3.2 $10^{-3}$  & 2.4 $10^{-5}$  & 1.1 $10^{-2}$ \\ 
B2~2116+26  & 3.2 $10^{-2}$  & 3.7 $10^{-6}$  & 1.0 $10^{-4}$ \\
B2~2236+35  & 1.9 $10^{-2}$  & 6.6 $10^{-7}$  & 2.7 $10^{-5}$ \\ 

\hline
                      
\end{tabular}
\end{center}
$^{\mathrm{a}}$ Calculated assuming parameters from the spectral
analysis (see text).  

{\bf Columns Explanation}: 1=Source name; 2=Bondi accretion rate ($M_{\odot}/{\rm yr}$);
3=Ratio of the bolometric luminosity to the Eddington luminosity;
4=Radiative efficiency.
No values for 3C~438 has been obtained. 
\label{accretion}
\end{table}

\end{document}